\documentclass[prd,11pt,onecolumn,tightenlines,nofootinbib,superscriptaddress]{revtex4}
\usepackage{amsmath,amsfonts}
\usepackage{subfigure}
\usepackage{enumerate}
\usepackage{braket}
\usepackage[colorlinks,citecolor=blue,linkcolor=blue,urlcolor=blue]{hyperref}
\usepackage{graphicx}

\newcommand{\M}{{\mathbb M}}
\newcommand{\R}{{\mathbb R}}
\newcommand{\N}{{\mathbb N}}
\newcommand{\va}{\scriptscriptstyle}
\newcommand{\heq}{\,\hat=\,}
\newcommand{\D}{\Delta}
\newcommand{\be}{\begin{equation}}
\newcommand{\ee}{\end{equation}}
\DeclareMathAlphabet{\mathfs}{U}{rsfs}{m}{n} 
\newcommand{\mfs}[1]{\mathfs {#1}}         
\newcommand{\s}{{\kappa_{\va SG}}}
\newcommand{\scri}{{\mfs J}}
\newcommand{\sL}{{\mfs L}}
\newcommand{\sI}{{\mfs I}}
\newcommand{\ba}{\nopagebreak[3]\begin{eqnarray}}
\newcommand{\ea}{\end{eqnarray}}

\begin{document}

\title{\Large\scshape\bfseries Light Cone Thermodynamics}

\author{Tommaso De Lorenzo}
\email{tommaso.de-lorenzo@cpt.univ-mrs.fr}
\affiliation{Aix Marseille Univ, Universit\'e de Toulon, CNRS, CPT, 13000 Marseille, France}
\author{Alejandro Perez}
\email{perez@cpt.univ-mrs.fr}
\affiliation{Aix Marseille Univ, Universit\'e de Toulon, CNRS, CPT, 13000 Marseille, France}
\affiliation{FaMAF, UNC; Instituto de F\'isica Enrique Gaviola (IFEG), CONICET, \\
Ciudad Universitaria, 5000 C\'ordoba, Argentina}
\begin{abstract}
We show that null surfaces defined by the outgoing and infalling wave fronts emanating from and arriving at a sphere in Minkowski spacetime have thermodynamical properties that are in strict formal correspondence with those of black hole horizons in curved spacetimes. Such null surfaces, made of pieces of light cones, are bifurcate conformal Killing horizons for suitable conformally stationary observers. They can be extremal and non-extremal depending on the radius of the shining sphere. Such conformal Killing horizons have a constant light cone (conformal) temperature, given by the standard expression in terms of the generalisation of surface gravity for conformal Killing horizons. Exchanges of conformally invariant energy across the horizon are described by a first law where entropy changes are given by $1/(4\ell_p^2)$ of the changes of a geometric quantity with the meaning of horizon area in a suitable conformal frame. These conformal horizons satisfy the zeroth to the third laws of thermodynamics in an appropriate way. In the extremal case they become light cones associated with a single event; these have vanishing temperature as well as vanishing entropy.
\end{abstract}

\maketitle

\section{Introduction: The Results in a Nutshell}

Classical Black Holes behave in analogy with thermodynamical systems \cite{Bardeen1973}. According to general relativity they satisfy the four laws of black hole mechanics. 
The surface gravity $\kappa_{\va SG}$ of a stationary black hole is constant on the horizon: {\em the zeroth law}. Under small  perturbations,  stationary black holes---which are characterized by a mass $M$, an angular momentum $J$,  and a charge $Q$---satisfy the {\em first law}
\be\label{1st}
\delta M=\frac{\kappa_{\va SG}}{8\pi} \delta A+\Omega \delta J+\Phi \delta Q, 
\ee 
where $\Omega$, $\Phi$, and $A$ are the angular velocity, electrostatic potential, and area of the stationary (Killing) horizon.  The Hawking area theorem \cite{hawking1973large}
\be\label{arealaw}
\Delta A\ge0
\ee
is regarded as the {\em second law}. The {\em third law}---expected to be valid from the cosmic censorship conjecture  \cite{Penrose:1969pc}---corresponds to the statement that extremal black holes, for which $\kappa_{\va SG}=0$, cannot be obtained from a non extremal one by a finite sequence of physical processes. 
When quantum effects are considered these analogies become facts of semiclassical gravity. Stationary black holes radiate particles in a thermal spectrum with temperature $T=\kappa_{\va SG}/(2\pi)$ \cite{hawking1974black,hawking1975} so that the first term in \eqref{1st} can be interpreted as a heat term expressed in terms of changes in the black hole entropy $S=A/4$ in Planck units. The area law \eqref{arealaw} is promoted to the generalized second law \cite{PhysRevD.7.2333}.  
In the quantum realm another statement that could be associated to the  third law is that extremal black holes should have vanishing entropy, an argument for which can be found in \cite{Hawking:1994ii}.

All this is naturally interpreted as providing valuable information about the quantum theory of gravity of which the semiclassical treatment should be a suitable limit of. In this respect it is useful to have examples of a similar behaviour in simplified situations. The Fulling-Davies-Unruh thermal properties associated to quantum field theory in flat spacetimes \cite{Fulling:1972md,Davies:1974th,Unruh:1976db} is often used as an example illustrating, in a simplified arena, some of the aspects behind the physics of black holes. In this respect, the Rindler (Killing) horizon associated with a family of constantly accelerated observers in Minkowski spacetime is taken as an analogue of the black hole horizon. This analogy is supported further by the 
statement that the near horizon geometry of a non-extremal black hole can be described, in suitable coordinates, by the metric
\be\label{nhl}
ds^2=-\kappa_{\va SG} R^2 dt^2+dR^2+(r_{\va H}^2 + a^2) dS^2+O(R^3)
\ee
where $dS^2 = d\vartheta^2 + \sin^2\vartheta d\varphi^2$ is the metric of the unit two-sphere, $a\equiv J/M$. At least in the $(R,t)$ ``plane'', the above equation matches the 2d Rindler metric where $R=0$ is the location of the black hole horizon. The thermodynamical properties of Rindler horizons have been extensively discussed in the literature \cite{PhysRevD.87.124031,PhysRevLett.75.1260,PhysRevD.82.124019,PhysRevD.82.024010,Padmanabhan:2009vy}. 
However, strictly speaking, the near horizon geometry is not Rindler due to the presence of the term $r_{H}^2 dS^2$ in the previous equation that makes the topology of the horizon $S^2\times \R$ instead of $\R^3$, which implies the area of the black hole horizon to be finite $A=4\pi (r^2_{H}+a^2)$ instead of infinite. Another obvious difference is that, in contrast with Rindler, the Riemann curvature is non zero at the black hole horizon. Only in the infinite area limit the local geometry becomes exactly that of a Rindler horizon. 
Furthermore, in contrast with black hole horizons, the Rindler horizon has a domain of dependence that includes the whole of what one would regard as the {\em outside} region. In fact the Rindler horizon defines  a good initial value characteristic surface.
More precisely, any regular initial data for an hyperbolic equation  such as a Klein-Gordon or Maxwell field with support on the corresponding wedge---what would be the {\em outside}---can be encoded in data on the Rindler horizon \cite{wald2010general}. An implication of this is that no energy flow can actually escape to infinity without crossing the Rindler horizon. No notion analogous to the asymptotic observers outside of the black hole exists when considering the Rindler wedge and its Killing horizon boundary. A geometric way to stating this is that the Rindler horizon is given by the union of the past light cone of a point at $\sI^+$ with the future light cone of a point at $\sI^-$. In this sense the Rindler wedge is better described as a limiting case of the interior of finite diamonds---see next paragraph---rather than representing faithfully the outside region of a black hole spacetime. 
In this work we show that there exists a more complete analogue of black holes in Minskowski spacetime. 

There is a natural interest in double cone regions in Minkowski spacetime, also called diamonds, in algebraic quantum field theory \cite{Hislop:1981uh} or in the link between entanglement entropy and Einstein equations \cite{PhysRevLett.116.201101}. The conformal relationship with the Rindler wedge has been used in order to define the corresponding modular Hamiltonian and study thermodynamical properties in \cite{Martinetti:2002sz, Martinetti:2008ja}.
Here we concentrate on the causal complement of the diamond, and show that it shares several analogies with the exterior region of a stationary spherically symmetric black hole.
  
Such flat spacetime regions as the diamond and its complement are directly related to the geometry of radial Minkowski  Conformal Killing vector Fields (MCKFs). What we shall show is that radial MCKFs can be classified in a natural correspondence with black holes spacetimes of the Reissner-Nordstrom (RN) family (i.e. $J=0$). They can be timelike everywhere in correspondence with the naked singularity case $M^2 < Q^2$ where the stationarity Killing field is timelike everywhere. But more interestingly, radial MCKFs can become null, and being surface forming, generate conformal Killing horizons. As we will show in Section \ref{MCKF} these are conformal bifurcate Killing horizons analogue, in a suitable sense, to the black hole horizons in the RN family. The results of this paper can be summarised as follows:

\begin{enumerate}
\item   \label{item} {\em Radial MCKFs define conformal Killing horizons:} Radial MCKFs become null on the light cones of two events on Minkowski spacetime that are separated by a timelike interval. By means of a Lorentz transformation these two events can be located on the time axis of an inertial frame. A further time translation can place the two events in a time reflection symmetric configuration so that the symmetry $t\to -t$ of Reissner-Nordstrom spacetimes is reproduced.

\item {\em They have the same topology as black hole Killing horizons:} The topology of the conformal Killing horizon in Minkowski spacetime is $S^2\times \R$ as for the Killing horizons of the RN spacetime. 

\item {\em These horizons are of the bifurcate type:} Radial MCKFs vanish on a 2-dimensional sphere of radius $r_H$ and finite area $A=4\pi r_H^2$ that is the analogue of the minimal surface where the Killing horizon of the RN black hole vanishes. The bifurcate surface is the intersection of the two light cones described above.

\item  {\em They separate events in spacetime as in the BH case:} The global structure of the radial MCKF is closely analogous to the one of the Killing horizon of the RN spacetime. 
More precisely, there are basically the same worth of regions where the radial MCKF and the RN time translational Killing vector field is timelike and spacelike respectively.  In the non-extremal case there are outer and inner horizons in correspondence to the non-extremal RN solution. One of the two asymptotically flat regions of the maximally extended RN spacetime corresponds to the points in the  domain of dependence of the portion of the $t=0$ hypersurface in Minkowski spacetime inside the bifurcate sphere, namely {\em the diamond}; the other asymptotically flat region corresponds to the domain of dependence of its causal complement, namely {\em the black hole} exterior in our analogy. There are regions where the radial MCKF becomes spacelike. These  too are in correspondence with regions in the non-extremal RN black hole, namely the regions between the inner and the outer horizons.  In the extremal limit the regions where the radial MCKF is spacelike, as well as one of the asymptotic region, disappear and the correspondence with the extremal RN solution is maintained.
All this will be shown in detail in the following section; the correspondence is illustrated in Figure~\ref{fig:penrose}.

\item {\em They satisfy the zeroth law:} The suitably generalized notion of surface gravity $\kappa_{\va SG}$ is constant on the conformal Killing horizon: {\em the zeroth law}.  Extremal Killing horizons have $\s=0$. 

\item     {\em They satisfy the first law:} Considering the effects of matter perturbations described by a conformally invariant matter model, one can show that radial conformal Killing horizons satisfy the balance law $\delta M=\kappa_{\va SG}\, \delta A/ (8 \pi)+\delta M_\infty$: \emph{the first law}. Here $\delta M$ is the conformally invariant mass of the perturbation, $\delta M_\infty$ is the amount of conformal energy flowing out to future null infinity $\sI^+$, and  $\delta A$ is what we call  {\em conformal area change}. The name stems from the fact that it corresponds to the change of a geometric notion with the meaning of horizon area in the appropriate conformal frame.

\item  {\em They satisfy the second law:} In the type of processes considered above and assuming the usual energy conditions, $\delta A\ge 0$: {\em the second law}. 

\item {\em They have constant (conformal) temperature:} When quantum fields are considered, a constant Hawking-like temperature $T=\kappa_{\va SG}/(2\pi)$ can be assigned to radial MCKFs. In view of this, $\delta S=\delta A/4$ in Planck units acquires the meaning of entropy variation of the conformal horizon.

\item {\em They satisfy a version of the third law:} Extremal radial MCKFs have vanishing temperature as well as vanishing entropy: {\em the third law}.

\item {\em Minkowski vacuum is the associated Hartle-Hawking state:} The Minkowski vacuum of any conformally invariant quantum field can be seen as the state of thermal equilibrium---usually called Hartle-Hawking state---in the Fock space defined with respect to the MCKF.

\item The near MCKF horizon limit matches the expression \eqref{nhl} with $a=0$. The Rindler horizon limit could be obtained by sending the events mentioned in Item \ref{item} suitably to future $i^+$ and past $i^-$ timelike infinity respectively. In that limit $r_{\va H}\to \infty$, the near horizon metric becomes the Rindler metric, and the corresponding radial MCKF becomes the familiar boost Killing field.
   
\end{enumerate}

The properties listed above will be discussed in more detail in the sections that follow. In Section \ref{MCKF} we construct general radial MCKFs
from the generators of the conformal group $SO(5,1)$, and explain their geometry. The causal domains they define and the analogy with black holes shown in Figure~\ref{fig:penrose} will be clarified there. The analogue of classical laws of black hole thermodynamics are shown to hold in a suitable sense for radial MCKFs in Section \ref{sec:LCthermo}. Finally we show, in Section \ref{HRCT}, that a semiclassical temperature can be assigned to radial MCKFs and we discuss the physical meaning of that temperature. 

\section{Conformal Killing Fields in Minkowski Spacetime}\label{MCKF}
\begin{figure}[!h]
\center
\begin{center}
\begin{minipage}[c]{.33\textwidth}
\centering
	\includegraphics[height=8cm]{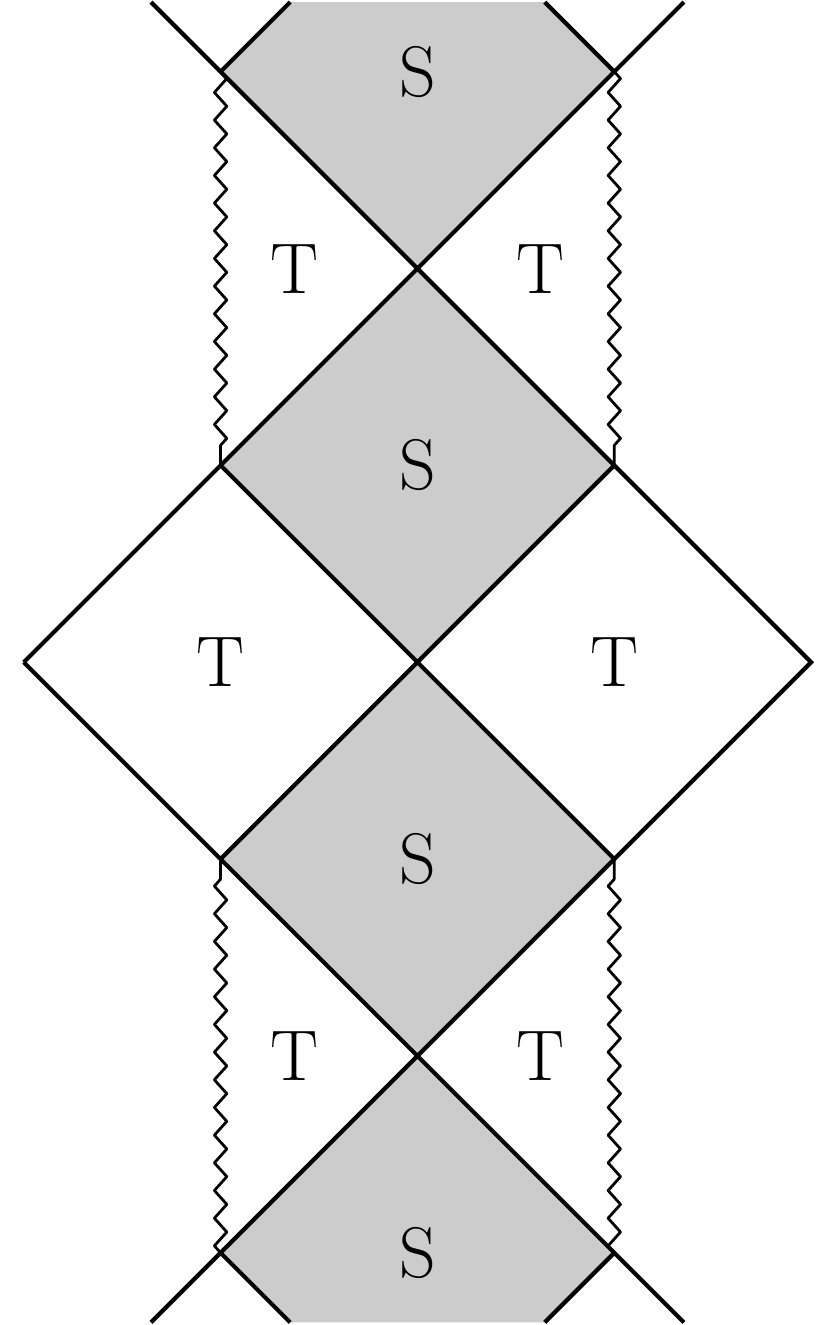}
\end{minipage}%
\begin{minipage}[c]{.6\textwidth}
\centering
          \includegraphics[height=8cm]{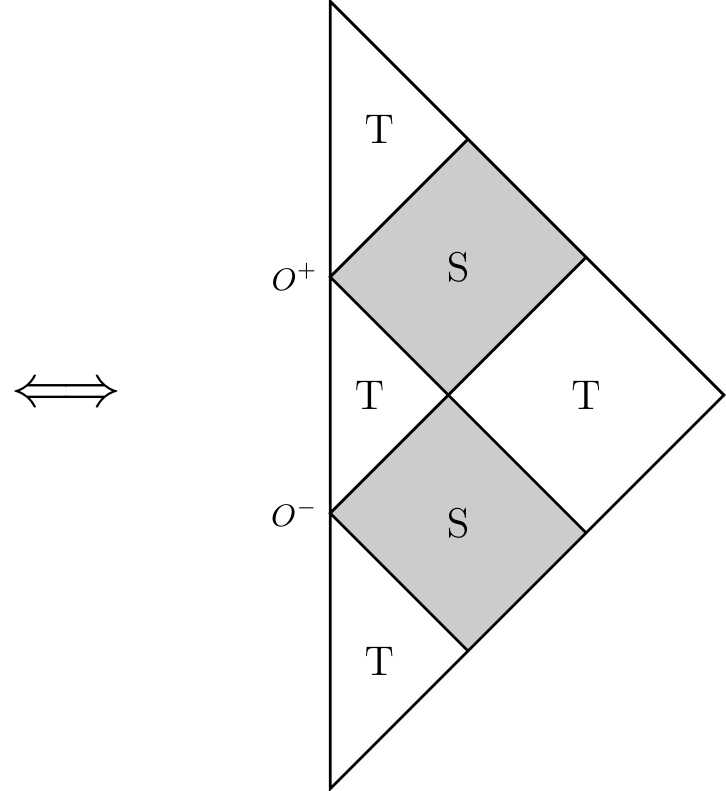}
\end{minipage}\\
\vspace{2cm}
\begin{minipage}[c]{.33\textwidth}
	\hspace{.6cm}
	\includegraphics[height=8cm]{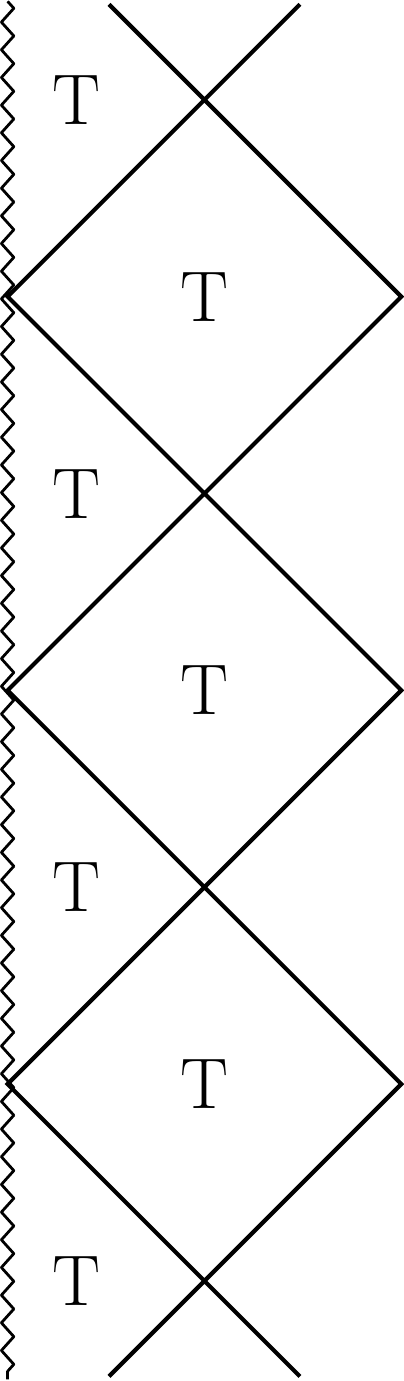}
\end{minipage}
\begin{minipage}[c]{.6\textwidth}
\centering
          \includegraphics[height=8cm]{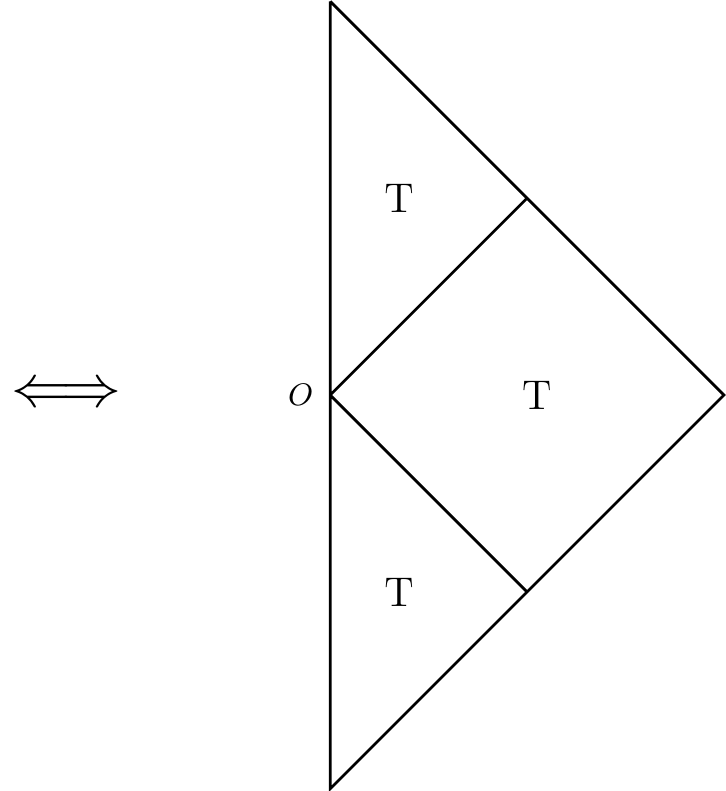}
\end{minipage}
\end{center}

\caption{The Penrose diagram of the Reissner-Nordstrom black hole on the left compared with the causal structure of the radial CKF in Minkowski spacetime on the right, in both the non-extremal $\Delta>0$ and extremal $\Delta =0$ case. The letters $S$ and $T$ designate the regions where the Killing or conformal Killing fields are spacelike or timelike respectively. The light cone emanating from the points $O^\pm$ (and $O$ in the extremal case) are the hypersurface where the MCKF is null.}
\label{fig:penrose}
\end{figure}

The conformal group in four dimensional Minkowski spacetime $\M^4$  is isomorphic to the group $SO(5,1)$ with
its 15 generators given explicitly by \cite{francesco2012conformal}
\be
\begin{split}
& P_\mu=\partial_{\mu} \ \ \ \ \ \ \ \ \ \ \ \ \ \ \ \ \ \ \ \ \ \ \ \ \ \ \ \ \ \ {\rm Translations} \\
& L_{\mu\nu}=\left(x_\nu\partial_{\mu} -x_\mu\partial_{\nu}\right)  \ \ \ \ \ \ \ \ \ \ \ \ \ {\rm Lorentz\ transformations} \\
& D= x^\mu\partial_{\mu}  \ \ \ \ \ \ \ \ \ \ \ \ \ \ \ \ \ \  \ \ \ \ \ \ \ \ \ \  {\rm Dilations}\\
& K_{\mu}=\left(2 x_\mu x^\nu \partial_{\nu} -x\cdot x\, \partial_{\mu}\right)\ \ \ \ \ \ \, {\rm Special\ conformal\ transformations,} 
\end{split}
\ee
where $f\cdot g \equiv f^\mu g_\mu$.
Any generator defines a Conformal Killing Field in Minkowski spacetime (MCKF), namely a vector field $\xi$ along which the metric $\eta_{ab}$ changes only by a conformal factor:
\be\label{lietext}
\sL_\xi \,\eta_{ab}=\nabla_a\xi_b+\nabla_b\xi_a=\frac{\psi}{2} \eta_{ab}
\ee
with
\be
\psi = \nabla_a \xi^a\,.
\ee
Consider now the Minkowski metric in spherical coordinates
\be\label{eq:min}
ds^2 = -dt^2 + dr^2 + r^2 dS^2\,.
\ee
Then dilations can be written as
\be
D=r\partial_r+t\partial_t
\ee
and $K_0$ as
\be
K_0=-2t D-(r^2-t^2) P_0\,.
\ee
Together with $P_0=\partial_t$, those are the only generators that do not contain angular components. Hence the most general radial MCKF has the form 
\be \label{explicit}
\begin{split}
\xi =&-a K_0+b D+c P_0\\
 =&(2at+b) D+ [a(r^2-t^2)+c]  P_0, 
\end{split}
\ee
with $a,b,c$ arbitrary constants.
Explicitly
\be\label{eq:CKF}
\begin{split}
\xi^\mu \partial_\mu &= \big[a(t^2+r^2)+bt+c\big]\partial_t + r(2at + b)\partial_r\\
&=(av^2+bv+c)\partial_v+(au^2+bu+c)\partial_u\,,
\end{split}
\ee
where $v=t+r$, $u=t-r$ are the standard null coordinates. The norm of $\xi$ is easily computed to be
\be
\label{eq:normabc}
\xi \cdot \xi= - (av^2+bv+c)(au^2+bu+c)\,.
\ee
Its causal behaviour, therefore, can be studied introducing the quantity
\be\label{Delta}
\Delta \equiv b^2 - 4ac\,.
\ee
The complete classification of such MCKFs is given in \cite{RCKF}. Here we are interested in the case $a \neq 0$, where we have three different types of behaviour depending on the sign of the parameter $\Delta$. 
When $\Delta < 0$, the MCKF is timelike everywhere, like the stationarity Killing field in the Reissner-Nordstrom solutions with a naked singularity $M^2<Q^2$. When $\Delta > 0$, on the other hand, the MCKF is null along two constant $u$ and two constant $v$ null hypersurfaces respectively given by 
\be
u_\pm = \frac{-b \pm \sqrt{\Delta}}{2a} \ \ \ \ {\rm or} \ \ \ \  v_\pm = \frac{-b \pm \sqrt{\Delta}}{2a}\,.
\ee
In other words, the MCKF is null on the past and future light cones of two points $O^{\pm}$, with coordinates given respectively by $O^{\pm} : (t = u_{\pm}, r = 0)$. These two light cones divide Minkowski spacetime into six regions. In those regions the norm of the MCKF changes going from timelike to spacelike as depicted on the top right panel of Figure~\ref{fig:penrose}. The boundary of these regions are null surfaces generated by the MCKF; they define conformal Killing horizons. The vector field $\xi$ vanishes at the bifurcate 2-dimensional surface defined by the intersection of the previous null surfaces, namely at the sphere
\be\label{rh}
t=t_H=-\frac{b}{2a}\,, \qquad r=r_H=\frac{\sqrt{\Delta}}{2a}\,.
\ee
The {\em ``extremal''} case $\Delta=0$ is a limiting case between the other two: the MCKF is null on the light cones $u_0=v_0=-b/(2a)$ emanating from a single point $O$, and timelike everywhere else. This is depicted in the bottom right panel of Figure~\ref{fig:penrose}.

It is interesting to notice that the four regions around the bifurcate sphere in the non-extremal case are in one-to-one correspondence with the corresponding four regions around the bifurcate sphere in the case of stationary black holes of the Reissner-Nordstrom family. The correspondence is maintained in the extremal limit where the bifurcate sphere degenerates to a point and the four regions collapse to a single one. In the black hole case the bifurcate sphere is pushed to infinity and one of the asymptotically flat regions disappears. In our case the bifurcate sphere is shrunk to a point at the origin and the region in the interior of the light cones, the diamond, disappears. The analogy is emphasised in Figure~\ref{fig:penrose}.

\begin{figure}[t]
\center
\subfigure[]{
   \includegraphics[height=9cm]{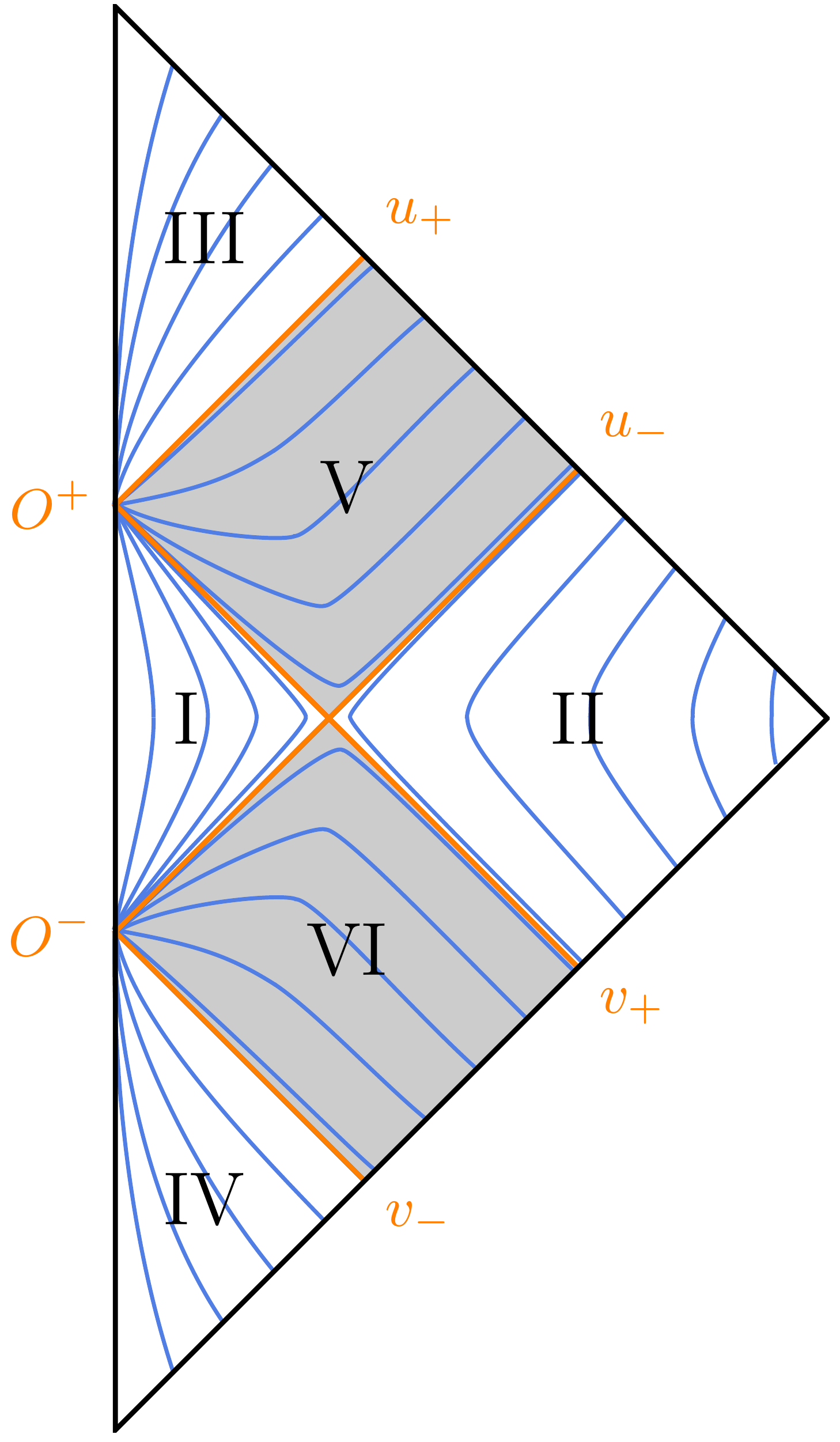}}
\hspace{15ex}
\subfigure[]{
   \includegraphics[height=9cm]{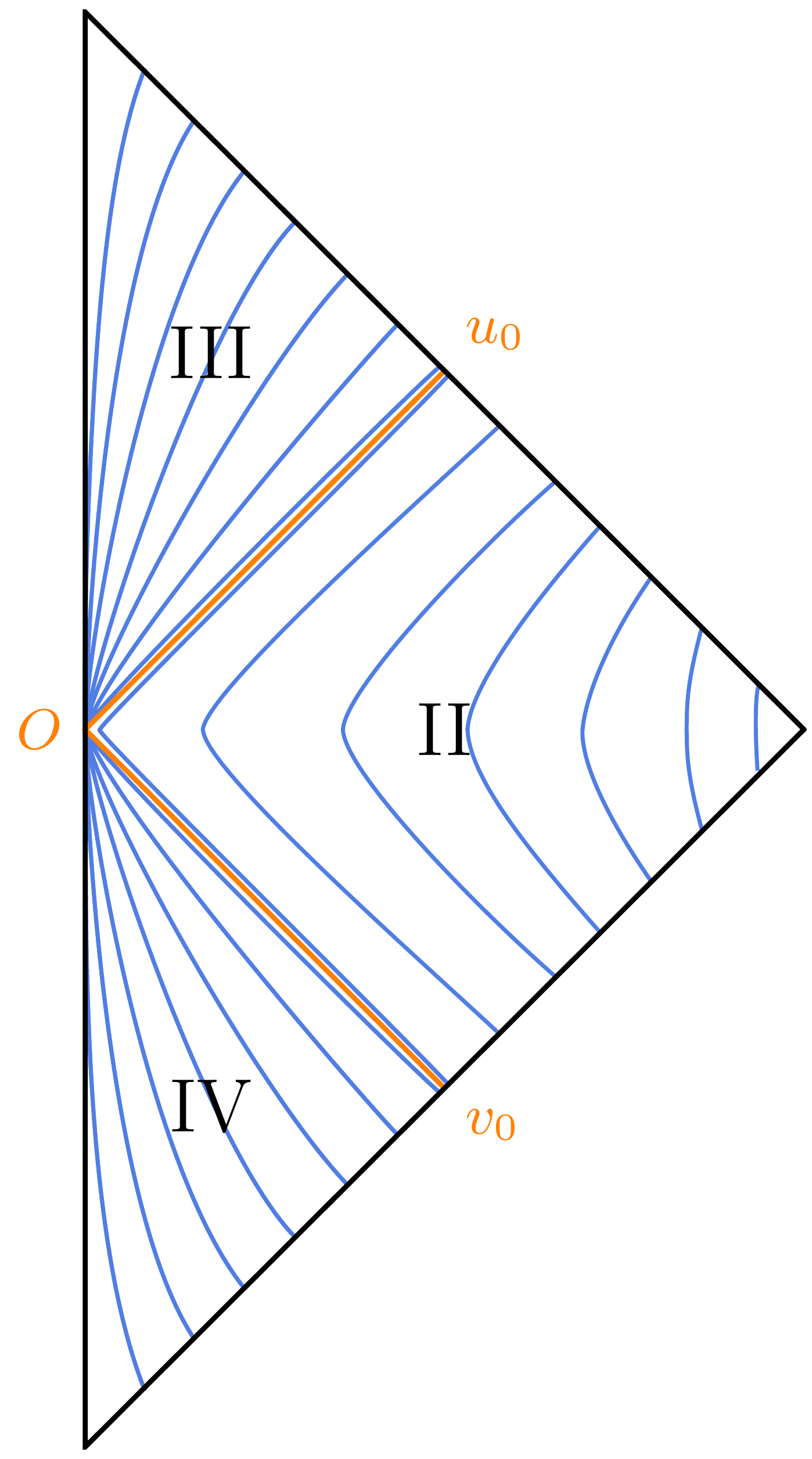}}
\caption{The flow of the radial MCKF, depending on the value of the parameter $\Delta$, Eq.~\eqref{Delta}. When $\Delta < 0$: the MCKF is timelike everywhere. a) $\Delta > 0$: the Minkowski spacetime is divided into 6 different regions, where the norm of the MCKF changes from being timelike to being spacelike, through being null along the four light rays $u_\pm$, $v_\pm$. b) $\Delta = 0$: the MCKF is everywhere timelike except for the two null rays $u_0=v_0$ where it is null.}
\label{fig:families}
\end{figure}

The flow of $\xi$ describes uniformly accelerated observers, with integral curves being a one parameter family of rectangular hyperbolas given by \cite{RCKF}
\be\label{eq:hyperbolas}
t^2 - \left(r+\frac{\zeta}{2a}\right)^2 = \frac{\Delta - \zeta^2}{4a^2}\,,
\ee
where $\zeta$ is the parameter labeling members of the family. The complete situation is depicted in Figure~\ref{fig:families}. From the picture it is clear that, seen from the point of view of the observers that follow the MCKF in Region II, the boundary of the region is a {\em bifurcate conformal Killing horizon} with topology $S^2 \times \R$. This is the same topology as the one of {\em bifurcate Killing horizons} of stationary black holes in the asymptotically flat spacetime context.
To summarise: radial conformal Killing fields in Minkowski spacetime generate bifurcate conformal Killing horizons that reproduce the main topological features of stationary spherically symmetric Killing horizons. This is the first obvious indication that makes MCKFs interesting for drawing analogies with black holes. The aim of what follows is to show that this analogy is more profound and extends very nicely to the thermodynamical properties of black hole Killing horizons. In Section~\ref{sec:LCthermo}, indeed, we will be able to define, in a suitable sense that will become clear, the four laws of thermodynamics for  bifurcate MCKF horizons.

\subsection{Introducing two geometric scales defining the radial MCKF} 

Here we associate the parameters $a, b$, and $c$ in \eqref{explicit} with geometric notions. First we set $b=0$ by means of a time translation $t\to t-b/(2a)$. In this way the points $O^{\pm}$ are placed on the time axis in the future and the past of the origin at equal timelike distance, and the distributions of regions become $t$-reflection symmetric. We make this choice from now on. Notice in addition that the parameter $a$ has dimension $[length]^{-2}$, while $c$ is dimensionless. We can therefore rewrite those constants in terms of two physical length scales. The first one is the radius of the bifurcate sphere $r=r_{\va H}$, Eq.~\eqref{rh}:
\be\label{rH}
r_{\va H}^2=\frac{\Delta}{4a^2} = -\frac{c}{a}\,.
\ee
There is also another natural geometric scale associated to the radius of the sphere $r_{\va O}$ at $t=0$ where we demand $\xi$ to be normalized. Such sphere represents the ensemble of events where the MCKF can be associated with the orbits of observers. We call this sphere the {\em observers sphere}. The normalization condition at $r_{\va O}$ is the analogue of the normalization condition for the stationarity Killing vector field at infinity in asymptotically flat  stationary spacetimes, e.g. stationary black holes, or the selection of a special observer trajectory when normalizing the boost Killing field in the Rindler wedge. Therefore, we demand the condition  $\xi\cdot\xi|_{t=0,r=r_{\va O}}=-1$ which, together with Eq.~\eqref{rH}, allows to determine both $a$ and $c$ as a function of $r_{\va H}$ and $r_{\va O}$. Explicitly one finds
\be
a = \frac{1}{r_{\va O}^2-r_{\va H}^2}\,, \ \ \ \ \ 
c = -\frac{r_{\va H}^2}{r_{\va O}^2-r_{\va H}^2}\,.
\ee
The radial conformal Killing field takes then the form
\be\label{eq:CKFb0}
\begin{split}
\xi^\mu \partial_\mu &= \frac{1}{r_{\va O}^2-r_{\va H}^2}\Big[(t^2+r^2-r_{\va H}^2)\partial_t + 2tr\,\partial_r\Big]\\
&=\frac{v^2-r_{\va H}^2}{r_{\va O}^2-r_{\va H}^2}\,\partial_v+\frac{u^2-r_{\va H}^2}{r_{\va O}^2-r_{\va H}^2}\,\partial_u\,;
\end{split}
\ee
its norm becomes
\be\label{eq:norm}
\xi \cdot \xi = - \frac{(v^2-r_{\va H}^2)(u^2-r_{\va H}^2)}{(r_{\va O}^2-r_{\va H}^2)^2}\,;
\ee
the parameter $\Delta$
\be
\Delta =\frac{4r_{\va H}^2}{(r_{\va O}^2-r_{\va H}^2)^2}\,,
\ee
which implies
\be
u_\pm = v_\pm = \pm r_{\va H}\,.
\ee
From equation \eqref{eq:CKFb0} we clearly see that $\xi$ vanishes at the bifurcate sphere $r=r_{\va H}$ and that $\xi=\partial_t$ at the observers sphere $r=r_{\va O}$; both spheres are defined to be on the $t=0$ surface. The vector field vanishes also at $O^{\pm}$. These two length scales completely determine the radial MCKF forming conformal Killing horizons.

\section{Light Cone Thermodynamics}\label{sec:LCthermo}
\begin{figure}[t]
\center
\includegraphics[height=9cm]{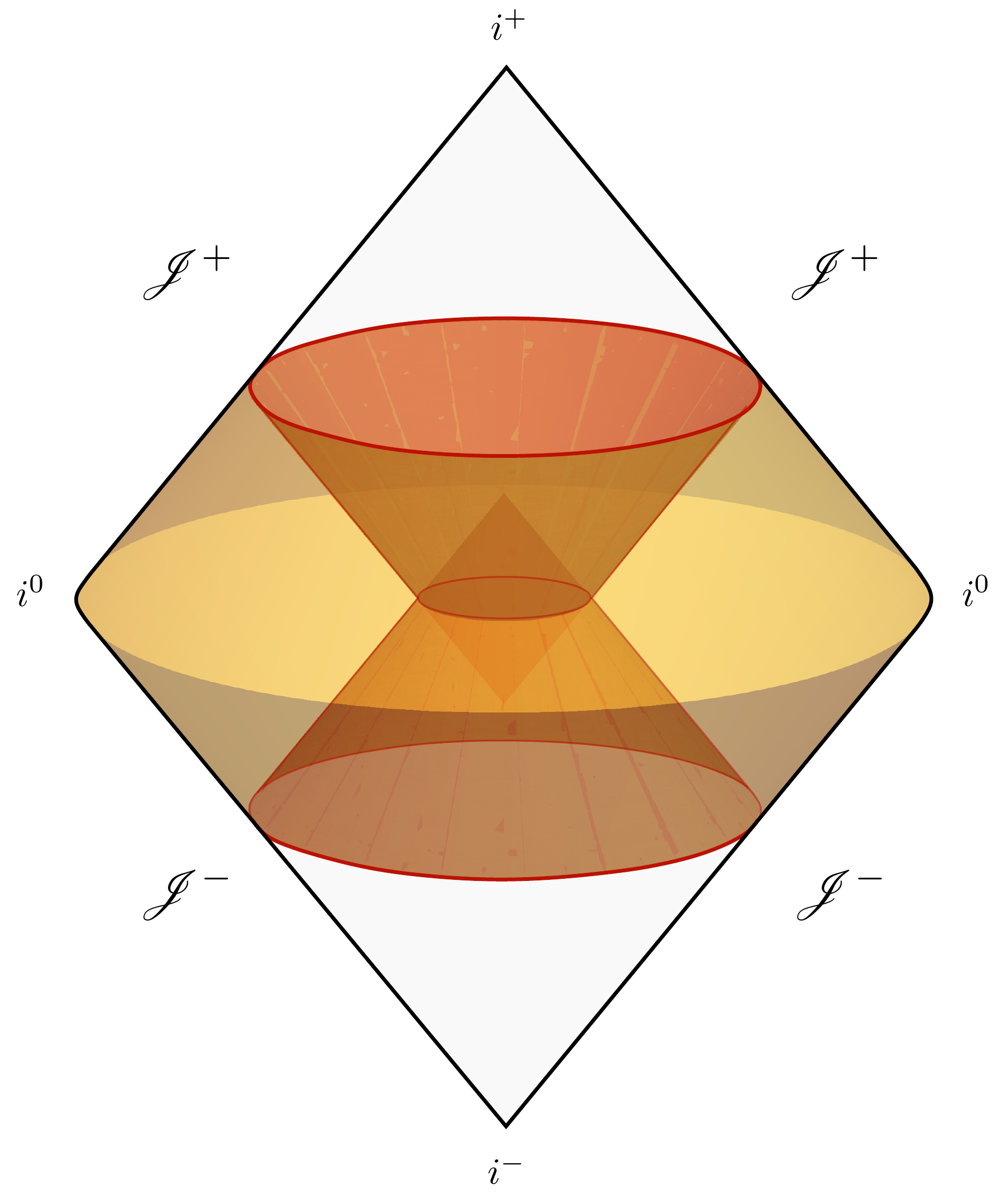} 
\caption{A $(2+1)$ dimensional diagram depicting the regions of interest in Minkowski spacetime. The two cones truncated at the sphere of radius $r=r_{\va H}$ represent the bifurcate conformal Killing horizon. They meet future and past null infinity on spherical cross-sections, represented by the two bigger rings. The central ring is the bifurcate sphere $r = r_H$, where $\xi=0$. The horizon is therefore a sphere with radius growing at the speed of light. In the center, one can also see the Cauchy development of the bifurcate sphere, \emph{the diamond}, Region I. Shaded in yellow is Region II, the region representing the outside of the horizon. Geometrically, it is the Cauchy development of the complement of Region I. The remaining part of Minkowski spacetime is occupied by Regions III to VI, which, to simplify the picture, are not clearly depicted here.}
\label{fig:Penrose_CD}
\end{figure}

In this central section of the paper, we will formulate the laws of thermodynamics for the bifurcate conformal Killing horizon generated by the radial MCKF. The horizon is defined by the two pieces of light cones meeting at the bifurcate sphere of radius $r_H$. It is the boundary of the causal complement of the diamond, Region II: the analogue of the exterior region of a stationary black hole spacetime. In Figure~\ref{fig:Penrose_CD}, the $S^2 \times \R$ topology of the horizon, together with the structure of Region II, is emphasised.

As already mentioned, see Eq.~\eqref{lietext}, a MCKF $\xi$ satisfies
\be
\sL_\xi \,\eta_{ab}=\nabla_a\xi_b+\nabla_b\xi_a=\frac{\psi}{2} \eta_{ab}
\ee
with
\be
\psi = \nabla_a \xi^a\,.
\ee
A conformal Killing horizon $H$ is defined as the surface where the MCKF is null, $\xi \cdot \xi=0$. Therefore, the gradient of $\xi\cdot\xi$ must be proportional to the normal to the horizon $\xi$. The proportionality factor defines the surface gravity $\s$ \footnote{\label{fofo} It is easy to check that $\s$ is invariant under conformal transformations $\eta_{ab}\to g_{ab}= \Omega^2 \eta_{ab}$ \cite{Jacobson:1993pf}. Under such transformations we have \be\nabla_a(g_{bc} \xi^b\xi^c)\hat=\nabla_a(\Omega^2 \eta_{bc} \xi^b\xi^c)\hat=\Omega^2 \nabla_a(\eta_{bc} \xi^b\xi^c)\hat=\s\Omega^2 \eta_{ab}\xi^b\hat=\s g_{ab}\xi^b,\ee 
where we used that $\xi\cdot \xi=0$ on the horizon. The surface gravity $\kappa_{\va SG}$ can be seen as the value of the acceleration of the conformal observers at the horizon when seen from the point of view of an observer placed at the observer sphere $r=r_{\va O}$ \cite{myThesis}.} via the equation
\be\label{susu}
\nabla_a (\xi\cdot \xi)\,\hat=-2\s \eta_{ab} \xi^b\,.
\ee
The symbol $\hat=$ stands for relations valid only at the horizon; we will use this notation whenever stressing such property is necessary. The CKF is also geodesic at the horizon \cite{sultana2004conformal,dyer1979conformal}, so that one can define the function $\kappa$ as
\be\label{eq:inaffinity}
\xi^a \nabla_a \xi^b {\,\hat=\,}\kappa\,\xi^b\,.
\ee
Thus $\kappa$ is the function measuring the failure of $\xi$ to be an affine geodesic on the horizon.
While for Killing horizons $\kappa = \s$, for conformal Killing horizons the following relation is valid:
\be
\s  = \kappa - \frac{\psi}{2}\,.
\ee
We will now use these relations in the special case of the MCKF defined in the previous sections.

\subsection{The zeroth law}
It is immediate to show that, for a general CKF, the quantity $\s$ is Lie dragged along the field itself \cite{sultana2004conformal,dyer1979conformal}; namely
\be
\sL_{\xi}\s = 0\,.
\ee
In our case, by spherical symmetry, this implies that $\s$ is actually constant on the horizon $H$ \footnote{It can however be shown for a general CKF under some assumptions \cite{sultana2004conformal}.}. This proves \emph{the zeroth law} of light cone thermodynamics.

\subsection{The first law}\label{sec:firstlaw}
Let us now define the energy-momentum current
\be\label{cucur}
J^a = T^{ab}\xi_b \,.
\ee
Conformally invariant field theories satisfy $T^a_{\;a}=0$ on shell \cite{francesco2012conformal}. For these theories, the current $J^a$ is conserved. Indeed
\be
\nabla_a J^a = \frac{\psi}{4} \;T^a_{\;a}=0 \,.
\ee
In such cases the current defines a conserved charge
\be\label{M}
M = \int_\Sigma T_{ab} \, \xi^a d\Sigma^b
\ee
with $d\Sigma^a$ being the volume element of a general Cauchy hypersuface $\Sigma$.

We want now to study the analogue of what is called the ``physical process version'' \cite{Hawking1972,Gao:2001ut} of the first law of black hole thermodynamics, which is also valid for more general bifurcate Killing horizons \cite{Amsel:2007mh} \footnote{There are works in the literature where the mechanical laws for conformal Killing horizons are investigated from the purely Hamiltonian perspective \cite{Chatterjee:2015fsa, Chatterjee:2014jda}. The strategy used in our specific and simple flat spacetime example seems more transparent for a deeper geometric insight.}. This will define the \emph{first law of light cone thermodynamics}. 
Let us therefore consider the process in which a small amount $\delta M$ of such ``energy'' \footnote{See Subsection~\ref{sec:energy} for a discussion on the meaning of such a conserved quantity.} passes through what plays the role of the future horizon $H^+$, namely the future light cone $u=u_-$. The passage of the matter will perturb the horizon. We will show that, at first order in linearised gravity, there is a balance law relating $\delta M$ with the conformal area change of the horizon $H^+$ (see below). 
The crucial difference in proving this relation with respect to the black hole case is that here the cross-sectional area of the horizon is changing along the affine geodesic generators of the light cone, even when no perturbation is considered. More technically, if we take the advanced time $v=t+r$ as an affine parameter along the null generators $\ell=\partial_v$ of $H^+$ in the flat background geometry, then one can show that the expansion $\theta$ is
\be\label{eq:MinExp}
\theta = \frac{1}{r} = \frac{2}{v-u}\,.
\ee
By definition, the expansion is the rate of change of the cross-sectional area with respect to the parameter \cite{poisson_2004}, namely,
\be\label{eq:expdS}
\theta =\frac{1}{dS} \frac{\partial dS}{\partial v}\,.
\ee
In our case $dS = 1/4\,(v-u)^2 \sin\vartheta d\vartheta d\varphi = $ and the two above equations are indeed in agreement.
The non vanishing of those quantities, therefore, implies that the area of the horizon is constantly increasing even if no flux of energy is considered. Consequently, we need to distinguish the two different changes in area: the background one that we call $dS_0$, and the one induced by the passage of the perturbation that we call $dS_1$. When a generic flux of energy is considered, we can write the expansion as
\be
\theta =\frac{1}{dS_0+dS_1} \frac{\partial}{\partial v} ( dS_0 + dS_1)\,.
\ee
In the approximation in which $dS_1 / dS_0  \ll 1$, we can expand it up to first order and find
\be\label{eq:ExpExp}
\begin{split}
\theta &= \frac{1}{dS_0}\frac{\partial dS_0}{\partial v} + \frac{1}{dS_0} \left(\frac{\partial dS_1}{\partial v} - \frac{1}{dS_0} \frac{\partial dS_0}{\partial v} dS_1\right)+ O\left( \frac{dS_1}{dS_0}\right)^2\\
&\equiv\theta_0 +\theta_1+ O\left( \frac{dS_1}{dS_0}\right)^2,
\end{split}
\ee
where $\theta_0$ is the unperturbed expansion given by Eqs.~\eqref{eq:MinExp}-\eqref{eq:expdS} and where we defined the perturbation 
\be\label{eq:th1}
\theta_1 \equiv \frac{1}{dS_0} \left(\frac{\partial dS_1}{\partial v} - \theta_0 dS_1\right)= \frac{\partial}{\partial v} \left(\frac{dS_1}{dS_0}\right)\,.
\ee
Moreover, by the Raychauduri equation one has that the variation of the expansion is connected with the flux of energy as \cite{wald2010general} 
\be\label{raycha}
\ell(\theta) = -\frac{1}{2} \theta^2 - \sigma_{ab}\sigma^{ab} + \omega_{ab}\omega^{ab} -8\pi T_{ab} \ell^a\ell^b\,;
\ee
where $\sigma_{ab}$ and $\omega_{ab}$ are the shear and twist tensors respectively, and $\ell(f) = \partial_v f$ for any function $f$.
Using Eq.~\eqref{eq:ExpExp} and the fact that, in this case $\omega^{0}_{ab}=\sigma^0_{ab}=0$, we get
\be
\ell(\theta_0) + \ell(\theta_1) = -\frac{1}{2} \theta_{0}{}^{2} -\theta_0\theta_1 -8\pi \delta T_{ab} \ell^a\ell^b+ O(\theta_1)^2\,,
\ee
where $\delta T_{ab}$ represents a small energy perturbation justifying the use of the perturbed equation \eqref{eq:ExpExp} \footnote{One may wonder whether $\ell$ is still an affinely parametrised generator of the horizon in the perturbed spacetime. This would be the case if the $\ell=du$ continues to be a null one-form for the metric $g_{ab}=\eta_{ab}+\delta g_{ab}$. Using the gauge symmetry of linearized gravity $\delta g_{ab}^\prime=\delta g_{ab}+2 \nabla^{\va (0)}_{(a}v_{b)}$ (for a vector field $v^a$), the condition $g^{ab} du_adu_b=0$ is equivalent to
 $\delta g^{uu}+\ell(\ell\cdot v)=0$, which can be solved for the gauge parameter $v$. }. 
Since $\theta_0$ is, by definition, the solution of the unperturbed Raychauduri equation---Eq.~\eqref{raycha} with $T_{ab} =0$---, the above equation reduces at first order to
\be\label{eq:rayTh1}
\ell(\theta_1) +\theta_0\theta_1 = - 8\pi \delta T_{ab} \ell^a\ell^b\,.
\ee
The last elements we need before starting the proof of the first law are the following: first notice from Eq.~\eqref{eq:CKFb0} that, on the horizon, our MCKF $\xi$ is parallel to $\ell$, namely
\be\label{alphaell}
\xi \heq \alpha \ell\,,
\ee
with $ \alpha = (v^2-r_{\va H}^2)/(r_{\va O}^2-r_{\va H}^2)$.
Substituting the above equation into definition \eqref{eq:inaffinity}, one can show that
\be\label{kappaell}
\kappa \,\hat=\, \ell(\alpha)=\partial_v \alpha \,.
\ee

Now we are ready to prove the first law of light cone thermodynamics. We consider a generic perturbation $\delta M$ of conformally invariant energy defined by \eqref{M}. Since this quantity is conserved, we can choose the Cauchy surface to integrate over in the more convenient way. For our purposes, we choose the union of the future horizon $H^+$ with the piece of $\sI^+$ contained in Region II, the latter being $\Sigma_\infty \equiv \rm {II} \cap \sI^+$. Eq.~\eqref{M} therefore becomes
\be\label{42}
\begin{split}
\delta M &=\int\limits_{H^+} \delta T_{ab} \, \xi^a d\Sigma^b_{0}+\int\limits_{\Sigma_\infty} \delta T_{ab} \xi^a d\Sigma_\infty^a=\\
 &= \int\limits_{H^+} \alpha \delta T_{ab} \,\ell^a\ell^b \,dS_0 dv +\delta M_{\infty}\\
 &= -\frac{1}{8\pi} \int\limits_{H^+} \alpha \Big( \partial_v(\theta_1) +\theta_0\theta_1 \Big) dS_0 dv+\delta M_{\infty} \\
 &= -\frac{1}{8\pi} \int\limits_{H^+} \alpha \Big( \partial_v(\theta_1) dS_0 +\partial_v (dS_0)\theta_1 \Big) dv+\delta M_{\infty} \\
 &= -\frac{1}{8\pi} \left[ \oint_{H^+(v)} \alpha\,\theta_1 dS_0 \Bigg{|}_{v_+}^{\infty} - \int\limits_{H^+} \kappa\, \theta_1 dS_0 dv \right]+\delta M_{\infty} \\
 &= \frac{\s}{8\pi} \int\limits_{H^+} \frac{\kappa}{\s} \,\theta_1 dS_0 dv+\delta M_{\infty}
\end{split}
\ee
To go from the first to the second line, we have used the fact that the unperturbed surface element of the horizon is given by $d\Sigma_0^a=\ell^a dS_0dv$, the fact that we defined the energy flux at infinity as
\be
\delta M_{\infty}\equiv\int\limits_{\Sigma_\infty} \delta T_{ab} \xi^a d\Sigma_\infty^b\,,
\ee
and the proportionality between $\xi$ and $\ell$, Eq.~\eqref{alphaell}. The third line follows from the perturbed Raychauduri equation \eqref{eq:rayTh1}. Using the definition of $\theta_0$ \eqref{eq:expdS} we obtain the fourth line. 
Integrating by parts and using (\ref{kappaell}) leads to line five. The boundary term vanishes from the fact that $\alpha(r_H)=0$, i.e. $\xi^a=0$ at the bifurcate surface, and that we choose initial condition at infinity in the usual teleological manner \cite{Gao:2001ut}, namely $\theta_1(\infty)=0$. 

We define the conformal area change $\delta A$ of the horizon as
\be \label{eq:area}
\delta A \equiv \int\limits_{H^+} \frac{\kappa}{\s} \,\theta_1 dS_0 dv= \frac{8\pi}{\s}{\int\limits_{H^+} \delta T_{ab} \,\xi^ad\Sigma^b},
\ee
in terms of which the first law follows
\be\label{firstLaw}
\delta M = \frac{\kappa_{\va SG}}{8\pi} \, {\delta A}+\delta M_{\infty}\,.
\ee

Some remarks are in order concerning the interpretation of equation \eqref{eq:area} and (\ref{firstLaw}):

First notice that the definition of $\delta A$ reduces to the standard expression when $\xi$ is a Killing field. Indeed, in that case $\kappa=\s$ and the unperturbed area of the horizon is constant, i.e. $\theta_0=0$.  From the definition \eqref{eq:th1}, \eqref{eq:area} is therefore the change in area of the Killing horizon. 

Now we argue that, in a suitable sense, \eqref{eq:area} retains the usual interpretation in the case of the expanding ($\theta_0\not=0$) conformal Killing horizon associated to a MKCF when perturbed with conformal invariant matter.
Under such circumstances, it can be verified that all the quantities appearing in the first law (\ref{firstLaw}) are conformal invariant; this follows directly from the conformal invariance of  $\s$ (see Footnote \ref{fofo}), and that of the flux density $ \delta T_{ab} \,\xi^a d\Sigma^b$ when conformal matter is considered \footnote{\label{FOOT}This a consequence of the fact that under a transformation $ g'_{ab}=\Omega^2 g_{ab}$ the energy momentum tensor of conformally invariant matter $\delta T_{ab}$ transforms as $\delta T'_{ab}=\Omega^{-2}\delta T_{ab}$ \cite{wald2010general}, and the volume element transforms as $d\Sigma'^a = \Omega^2 d\Sigma^a$, see Eq.~\eqref{sigmas}.}. Therefore, $\delta A$ is conformally invariant and this is the key for its geometric interpretation. To see this one can conformally map Minkowski spacetime to a new spacetime $g_{ab}=\Omega^2 \eta_{ab}$ where $\xi$ becomes a {\em bonafide} Killing vector field. Under such conformal transformation $\kappa \to \kappa_{\va SG}$ and $\theta_0 \to 0$, and thus the conformal invariant quantity $\delta A$ acquires the standard meaning of horizon area change, thus justifying its name. 
We show an explicit realization of such conformal map in Appendix \ref{AP2}.

Finally, as $\xi$ diverges at $\sI^+$ one might be worried that $\delta M_{\infty}$ might be divergent.
However, for massless fields the peeling properties of $T_{ab}$ are just the right ones for $\delta M_{\infty}$ to be convergent. Indeed this follows from the fact that $$T_{uu}=\frac{T_{uu}^0}{v^2}+O(v^{-3})\ \ \ \ \ \ \ T_{uv}=\frac{T_{uv}^0}{v^4}+O(v^{-5})\,,$$ and the form of $\xi$ given in \eqref{eq:CKFb0}. For a massless scalar field this is shown in \cite{wald2010general}; 
for Maxwell fields this can be seen in \cite{Adamo2012}. All this is expected from the fact that the current \eqref{cucur}
is conserved.

\subsection{The second law}

The quantity $\delta A$ is strictly positive in the context of first order perturbations of Minkowski spacetimes. 
This follows directly from the first law and the assumption that the conformal matter satisfies the energy condition 
$T_{ab} \ell^a\ell^b\ge 0$. It is the standard manifestation of the attractive nature of gravity in its linearized form. Needless is to say that this version of the second law is somewhat trivial in comparison with the very general area theorem for black hole \cite{hawking1973large}, as well as generic Killing \cite{chrusiel2001regularity}, horizons.

\subsection{The third law}

In our context the third law is valid in a very concrete and strict fashion. In the limit of extremality, $r_H\to 0$, the surface gravity $\kappa_{\va SG}\to 0$ and, the analogue of the entropy, the area $A$, goes to zero as well. This version of the third law is the analogue of the statement that at zero temperature the entropy vanishes, which is only true for systems with non degenerate ground states. No dynamical-process version of the third law  appears to make sense in our context. This might resonate at first sight with the statement \cite{Hawking:1994ii} that extremal BHs must have vanishing entropy, but the similarity is only in appearance as the area of the bifurcate sphere remains non-vanishing in the BH case.

\section{Quantum Effects: ``Hawking radiation'' and  Conformal Temperature} \label{HRCT}
When the laws of black hole mechanics where discovered, they were thought as a mere analogy with the one of thermodynamics. It is only after Hawking's discovery of semiclassical radiation \cite{hawking1974black,hawking1975} that they assumed a proper status of laws of black hole thermodynamics. In the previous Section, we established  the equivalent of the early analogy for the case of light cones in Minkowski spacetime and their gravitational perturbation. In what follows, we show that, also in this case, a semiclassical computation can be performed to give a thermodynamical meaning to those laws. In a suitable sense, radial MCKFs can be assigned a temperature
\be\label{temperature}
T = \frac{\s}{2\pi}\,.
\ee 
Thus the first law \eqref{firstLaw} becomes
\be
\delta M = T\delta S+\delta M_{\infty}\,.
\ee
with
\be
T = \frac{\s}{2\pi} \ \ \ \ \ \ {\rm and} \ \ \ \ \ \ \delta S= \frac{\delta A}{4}\,,
\ee
exactly as for stationary black holes.

To do so, let us start by noticing that for each region Minkowski spacetime is divided into by our radial MCKF $\xi$, there exists a coordinate transformation $(t,r,\vartheta,\varphi) \to (\tau,\rho,\vartheta,\varphi)$ adapted to the MCKF in the sense that $\xi(\tau)=-1$. The explicit maps are written in Appendix~\ref{app:coordinate}. Here we report the one for the region of interest, namely Region II. It reads \cite{HalILQGS}
\be
\label{eq:cootran}
\begin{split}
t &=\frac{\sqrt{\Delta}}{2a} \frac{\sinh(\tau \sqrt{\Delta})}{\cosh(\rho \sqrt{\Delta}) - \cosh(\tau \sqrt{\Delta})}\\
r &=\frac{\sqrt{\Delta}}{2a} \frac{\sinh(\rho \sqrt{\Delta})}{\cosh(\rho \sqrt{\Delta}) - \cosh(\tau \sqrt{\Delta})}\,,
\end{split}
\ee 
with $0\leq \rho < +\infty$ and $|\tau|<\rho$. Defining the null coordinates $\bar{v} = \tau + \rho$ and $\bar{u}=\tau - \rho$, the following relation with Minkowskian $u$ and $v$ is valid \cite{PhysRevD.26.1881}:
\be\label{uv}
\begin{split}
v &= t+r = -\frac{\sqrt{\Delta}}{2a} \coth\frac{\bar u\sqrt{\Delta}}{2}\\ 
u &= t-r = -\frac{\sqrt{\Delta}}{2a} \coth\frac{\bar v\sqrt{\Delta}}{2}\, ,
\end{split}
\ee
where, given the above mentioned restrictions on the coordinate, we have $\bar{u} \in (-\infty,0)$ and $\bar{v} \in (0,+\infty)$. The Minkowski metric \eqref{eq:min} becomes
\be \label{eq:confmetric}
ds^2 = \Omega^2 \left(-d\tau^2 + d\rho^2 +\Delta^{-1} \sinh^2(\rho \sqrt{\Delta})dS^2\right)
\ee
where the conformal factor takes the value
\be\label{eq:confFac}
\Omega = \frac{\Delta/2a}{\cosh(\rho \sqrt{\Delta})-\cosh(\tau \sqrt{\Delta})}\,.
\ee
As anticipated, the metric depends on the coordinate $\tau$ only through the conformal factor. The vector $\partial_\tau$, therefore, is a conformal Killing field for the Minkowski spacetime which can be shown to coincide with Eq.~\eqref{eq:CKFb0}. Explicitly
\be\label{eq:CKFtau}
\begin{split}
\xi^a \partial_a = \partial_\tau &= \left(a v^2 -\frac{\Delta}{4a}\right)\partial_v +  \left(a u^2 -\frac{\Delta}{4a}\right)\partial_u \\
&= (av^2 + c)\, \partial_v + (au^2 + c)\, \partial_u\\
&= \big(a(t^2+r^2)+c\big)\partial_t + 2art \,\partial_r\,.
\end{split}
\ee

\subsection{Bogoliubov transformations}

Consider now a scalar field $\phi$ evolving in Minkowski space. We will  define a vacuum state $\ket{0}$ and its corresponding Fock space $\mathcal{F}$ using the notion of positive frequency compatible with the notion of energy entering the first law Eq.~\eqref{firstLaw}. 
Making more precise what we anticipated in the first lines of this section, the remarkable result is that the standard Minkowski vacuum state is seen, in the Fock space $\mathcal{F}$, as a thermal state at the constant conformal temperature
\be
T = \frac{\s}{2\pi}\,.
\ee
The term conformal temperature is used because the state of the radiation looks thermal in terms of time translation notion associated to the conformal Killing time. See Section~\ref{sec:energy} for a detailed discussion, where the relationship between this notion of temperature and the physical temperature measured by a thermometer is also addressed.

Let us start by defining a meaningful notion of Fock space related to our conformally static observers. As for the discussion in the previous sections, Eq.~\eqref{firstLaw} holds only for conformally invariant matter models. For concreteness, here we consider a conformally invariant scalar field $\phi$ satisfying the conformally coupled Klein-Gordon (KG) equation  
\be\label{eq:KG}
\left(\Box^2-\frac16 R\right) \,\phi = 0\,,
\ee
where $\Box = g_{ab}\nabla^a\nabla^b$, with $\nabla^a$ the covariant derivative with respect to a general metric $g_{ab}$, and $R$ the  Ricci curvature scalar. The previous equation is conformally invariant in the sense that under a conformal transformation $g_{ab} \to g'_{ab} = C^2 g_{ab}$ solutions of \eqref{eq:KG} defined in terms of $g_{ab}$ are mapped into solutions of the same equation in terms of $g'_{ab}$ by the rule $\phi \to \phi'=C^{-1}\phi$ \cite{wald2010general,birrelldavies:QFTCST}.
 
As Eq.~\eqref{eq:confmetric} shows, the complement of the diamond in Minkowski space is conformally related to a region of a static Friedmann-Robertson-Walker (FRW) spacetime with negative spatial curvature $k=-|\Delta|$; see Appendix~\ref{app:FRW} for further details. The FRW Killing field $\partial_\tau$ corresponds to the Minkowski conformal Killing field in Eq.~\eqref{eq:CKFtau}. The strategy is therefore to find a complete set of solutions $U_i(x)$ of the KG equation in the static FRW spacetime, to deduce the one in our region using conformal invariance; here $i$ is a generic index labeling the modes. The Klein-Gordon equation \eqref{eq:KG} in the FRW spacetime under consideration reads
\be\label{eq:KG-exp}
\begin{split}
0 &= \left( \Box^2 -\frac16 R \right) U_i(x)\\
&= \left[ \frac{1}{\sqrt{-g}}\partial_\mu \left(\sqrt{-g}\partial^\mu \right) -\frac16 R \right] U_i(x)\\
&= \left[ -\partial_\tau^2 + \frac{1}{\sinh^2 (\rho\sqrt{\Delta})} \left( \partial_\rho \sinh^2(\rho\sqrt{\D})\, \partial_\rho + \frac{\Delta}{\sin\vartheta} \partial_\vartheta \sin\vartheta\, \partial_\vartheta + \frac{\D}{\sin^2\vartheta}\partial^2_\varphi \right) + \D\right] U_i(x) \,,
\end{split}
\ee
where $g = \det g_{ab}$ and we have used the fact that $R = -6 \D$. One can solve the previous equation by the ansatz 
\be
U^{\ell m}_\omega(x)=  \exp(- i \omega \tau) \frac{R_{\leftrightarrow \omega}^{\ell}(\rho)}{\sinh (\rho\sqrt{\Delta})} Y^{\ell m}(\vartheta, \varphi)\,,
\ee
which after  substitution  in  \eqref{eq:KG-exp}  gives
\be\label{eq:KGrho}
\left( \partial^2_\rho + \omega^2 -\frac{\ell(\ell+1) \, \D}{\sinh^2 (\rho\sqrt{\Delta})}\right) R^\ell_{\leftrightarrow \omega}(\rho) =0 \,,
\ee
where $\leftrightarrow$ denotes the two possible solutions: out-going modes will be denoted by a right arrow ($\rightarrow$) while in-going modes by a left arrow ($\leftarrow$).
Notice that we have substituted the generic index $i$ with the more specific $\omega$, $\ell$ and $m$. A complete set of solutions to this equation is given in \cite{birrelldavies:QFTCST}. These modes are  positive frequency modes with respect to the notion of time translation defined by the Killing time $\tau$, and they are orthonormal with respect to the Klein-Gordon scalar product, namely
\be
\begin{split}
( U_{\leftrightarrow \omega'}^{\ell' m'},U_{\leftrightarrow\omega}^{\ell m})  &= -i \int_\Sigma \left( U_{\leftrightarrow \omega}^{\ell m} \partial_a \bar{U}_{\leftrightarrow\omega'}^{\ell' m'} -\bar{U}_{\leftrightarrow\omega'}^{\ell' m'}\partial_a U_{\leftrightarrow\omega}^{\ell m}\right) d\Sigma^a \\
&= \delta_{\leftrightarrow}\delta^{\ell\ell'} \, \delta^{mm'} \, \delta(\omega,\omega')\,,
\end{split}
\ee
where $\delta_{\leftrightarrow}$ means that outgoing modes are orthogonal to ingoing ones.
As said at the beginning of the subsection, due to conformal invariance the set of modes \footnote{Such modes are the ``sphere modes'' considered in \cite{HalILQGS}.} defined by
\be\label{eq:modes}
u_{\leftrightarrow\omega}^{\ell m}(x) = \Omega^{-1}(x) U_{\leftrightarrow\omega}^{\ell m}(x) = \Omega^{-1}(x) \,e^{-i\omega \tau}  \frac{R_{\leftrightarrow\omega}^{\ell}(\rho)}{\sinh (\rho\sqrt{\Delta})} Y^{\ell m}(\vartheta, \varphi)
\ee
with $\Omega(x)$ given by Eq.~\eqref{eq:confFac}, are a complete set of solutions of the Klein-Gordon equation in our region of interest, the complement of the diamond in Minkowski space. Moreover, they satisfy
\be\label{eq:orto}
\begin{split}
(u_{\leftrightarrow\omega'}^{\ell' m'}, u_{\leftrightarrow\omega}^{\ell m}) &= (\Omega^{-1} \, U_{\leftrightarrow\omega}^{\ell m},\Omega^{-1} \,U_{\leftrightarrow\omega'}^{\ell' m'}) \\
&= -i \int_\Sigma \left[ \Omega^{-1}\, U_{\leftrightarrow\omega}^{\ell m} \partial_a \Big(\Omega^{-1} \bar{U}_{\leftrightarrow\omega'}^{\ell' m'}\Big) -\Omega^{-1} \bar{U}_{\leftrightarrow\omega'}^{\ell' m'}\partial_a \Big(\Omega^{-1}\,U_{\leftrightarrow\omega}^{\ell m} \Big)\right] d\Sigma_{\rm\va II}^a \\
&= -i \int_\Sigma \Big[ \Omega^{-2} \Big(U_{\leftrightarrow\omega}^{\ell m} \partial_a \bar{U}_{\leftrightarrow\omega'}^{\ell' m'} -\bar{U}_{\leftrightarrow\omega'}^{\ell' m'}\partial_a \,U_{\leftrightarrow\omega}^{\ell m}\Big) \\
&\phantom{= -i \int_\Sigma \Big[ \Omega^{-2} \Big(U_{\leftrightarrow\omega}^{\ell m} \partial_a \bar{U}_{\leftrightarrow\omega'}^{\ell' m'} -\bar{U}_{\leftrightarrow\omega'}^{\ell' m'}}+ \Omega^{-1} \bar{U}_{\leftrightarrow\omega'}^{\ell' m'} U_{\leftrightarrow\omega}^{\ell m} \Big(\partial_a \Omega^{-1} - \partial_a \Omega^{-1}\Big)\Big] d\Sigma_{\rm\va II}^a\\
&= -i \int_\Sigma \left( U_{\leftrightarrow\omega}^{\ell m} \partial_a \bar{U}_{\leftrightarrow\omega'}^{\ell' m'} -\bar{U}_{\leftrightarrow\omega'}^{\ell' m'}\partial_a \,U_{\leftrightarrow\omega}^{\ell m}\right) d\Sigma^a\\
&= (U_{\leftrightarrow\omega'}^{\ell' m'},U_{\leftrightarrow\omega}^{\ell m}) \,.
\end{split}
\ee
Here $\Sigma$ is a Cauchy surface shared by the two conformally related spacetimes; $d\Sigma^a$ and $d\Sigma_{\rm\va II}^a$ are the volume elements of $\Sigma$ in the static FRW spacetime and in the complement of the diamond respectively. The above result is given by the fact that the two volume elements are related by \footnote{This is generically true for any hypersurface $\Sigma$ shared between two conformally related spacetimes $g'_{ab}=C^{2} g_{ab}$. Indeed, if $n^a$ is the unit normal to $\Sigma$ with respect to $g_{ab}$, then $n'^a=C^{-1} n^a$ is the unit normal to $\Sigma$ with respect to $g'_{ab}$. The 3-dimensional volume elements, at the same time, are related by $\sqrt{h'}=C^{3} \sqrt{h}$. It follows that $d\Sigma'^a=C^2 d\Sigma^a$.}
\be\label{sigmas}
d\Sigma^a_{\rm\va II} = n_{\rm\va II}^a \sqrt{-h_{\rm\va II}} \, d^3 y = \Omega^2 n^a \sqrt{h} \,d^3y = \Omega^2 d\Sigma^a\,,
\ee
where $n^a$ is the normal to $\Sigma$, $h$ is the determinant of the intrinsic metric defining $\Sigma$ it self, and $y^i$ are the coordinate describing the latter. The subscript $\rm II$ indicates objects defined in Region II of Minkowski spacetime; the same objects without any subscript are in FRW. Eq.~\eqref{eq:orto} shows that the modes $u^{\ell m}_\omega$ provide a complete set of solutions inducing a positive definite scalar product, namely everything one needs to perform the standard quantisation procedure. Hence,
one can write the field operator in Region II of Minkowski spacetime as
\be
\begin{split}
\phi(x) &= \int_0^{+\infty} d\omega \sum_{\ell m}  \Big(a_{\leftarrow \omega}^{\ell m} u_{\leftarrow \omega}^{\ell m}(x) + a_{\leftarrow \omega}^{\ell m\,\dagger} {\bar u}_{\leftarrow \omega}^{\ell m}(x)\Big)+\Big(a_{\rightarrow \omega}^{\ell m} u_{\rightarrow \omega}^{\ell m}(x) + a_{\rightarrow \omega}^{\ell m\,\dagger} {\bar u}_{\rightarrow \omega}^{\ell m}(x)\Big) \\
 &= \int_0^{+\infty} d\omega\  \Omega^{-1}(x) \Big[\sum_{\ell m} \Big(a_{\leftarrow \omega}^{\ell m} U_{\leftarrow \omega}^{\ell m}(x) + a_{\leftarrow \omega}^{\ell m\,\dagger} U_{\leftarrow \omega}^{\ell m}(x)^*\Big) \\
 &\phantom{= \int_0^{+\infty} d\omega\  \Omega^{-1}(x) \Big[\sum_{\ell m} \Big(a_{\leftarrow \omega}^{\ell m} U_{\leftarrow \omega}^{\ell m}(x) + a_{\leftarrow \omega}^{\ell m\,\dagger}} + \Big(a_{\rightarrow \omega}^{\ell m} U_{\rightarrow \omega}^{\ell m}(x)+ a_{\rightarrow \omega}^{\ell m\,\dagger}{\bar U}_{\rightarrow \omega}^{\ell m}(x)\Big) \Big] \,,
 \end{split}
\ee
where $a_{\leftrightarrow \omega}^{\ell m \dagger}$ and $a_{\leftrightarrow \omega}^{\ell m}$ denote the creation and annihilation operators in the corresponding modes. The vacuum state $\ket{0}$ defined by $a_{\leftrightarrow \omega}^{\ell m} \ket{0}=0$ is usually called the  \emph{conformal vacuum} \cite{birrelldavies:QFTCST}. This state is highly pathological from the perspective of inertial observers. Indeed, it has vanishing entanglement with the interior of the diamond and would lead to a divergent energy momentum tensor at $H^+$. More precisely, this is not a Hadamard state. The same thing happens when one considers the Rindler vacuum defined by the boost Killing field.  

Let us notice now that Eq.~\eqref{eq:KGrho} is simplified in the limit $\rho \to +\infty$. This limit corresponds, in Region II and for $\tau > 0$, to the limit $\bar{v} \to +\infty$ or more clearly $u \to u_-$ with $v$ free to span the whole range $[v_+,+\infty)$. That is to say a ``near horizon limit'' \footnote{In the bottom part of our region, $\tau < 0$, this limit corresponds to $v \to v_+$, while $u$ free to vary. That is to say a near past horizon limit.}. 
In this limit the last term of Eq.~\eqref{eq:KGrho}, the one dependent on $\ell$, can be neglected. Solutions $R(\rho)$, therefore, do not depend on $\ell$ in such near horizon approximation and are simply given by $\exp(\pm i\omega \rho)$. The modes \eqref{eq:modes}, consequently, behave as
\be\label{modes}
{ u^{\ell m }_{\leftarrow \omega }}(x) +{ u^{\ell m }_{\rightarrow \omega }}(x) \approx  \frac{1}{ \sqrt{\omega}}\, \frac{e^{-i\omega \bar{u}} + e^{-i\omega \bar{v}}}{\Omega \, \sinh(\rho\sqrt\D)}Y^{\ell m}(\vartheta, \varphi)
 = \frac{\sqrt\D}{\sqrt{\omega}}\, \frac{e^{-i\omega \bar{u}} + e^{-i\omega \bar{v}}}{r}Y^{\ell m}(\vartheta, \varphi)
\,,
\ee
where $r$ is the Minkowskian radial coordinate and we have used definition \eqref{eq:cootran}.

Clearly, the solution of the Klein-Gordon equation and the consequent quantisation of the field can be carried out also in the whole Minkowski spacetime by considering inertial $r= const$ observers. This defines positive frequency modes $u^M_\omega$ with respect to the Killing field $\partial_t$, as well as a decomposition of the field as
\be
\phi(x) = \int_0^{+\infty} d\omega \Big({b^{\ell m}_{\leftarrow \omega }} {u^{\ell m M}_{\leftarrow \omega }}(x) + {b_{\leftarrow \omega }^{\ell m \dagger}}\, {\bar u^{\ell m M}_{\leftarrow \omega }}(x)\Big)+\Big({b^{\ell m}_{\rightarrow \omega }} {u^{\ell m M}_{\rightarrow \omega }}(x) + {b_{\rightarrow \omega }^{\ell m \dagger}}\, {\bar u^{\ell m M}_{\rightarrow \omega }}(x)\Big)\,.
\ee
In the limit $r\to +\infty$ the Minkowskian solutions can be approximated by 
\be\label{Mmodes}
{ u^{\ell m M}_{\leftarrow \omega }}(x) +{ u^{\ell m M}_{\rightarrow \omega }}(x) \approx   \frac{1}{\sqrt{\omega}}\, \frac{e^{-i\omega u} + e^{i\omega v}}{r}Y^{\ell m}(\vartheta, \varphi) \,.
\ee 
The standard Minkowski vacuum state $\ket{0}_M$ of the Fock space is defined by $b^{\ell m}_{\leftrightarrow \omega}\ket{0}_M=0$. The Minkowski modes are also orthonormal with respect to the Klein-Gordon scalar product, namely
\be\label{eq:ortoM}
(u^{\ell m M}_{\leftrightarrow \omega},u^{\ell' m' M}_{\leftrightarrow \omega'}) = \delta_{{\leftrightarrow}}\delta_{\ell\ell'}\delta_{mm'}\delta(\omega,\omega')\,,
\ee
which is immediately verified for outgoing and infalling modes by integrating on $\sI^+$ and $\sI^-$ solutions in the form \eqref{Mmodes}. 
The two different vacua are in general non-equivalent and one vacuum state can be a highly exited state in the Fock space defined by the other, and viceversa. This idea is formalised by introducing the so-called Bogoliubov transformations between the two complete sets of modes $u^{\ell m}_{\leftrightarrow \omega}$ and $u^{\ell m M}_{\leftrightarrow \omega}$. Briefly---for more details see for example \cite{wald2010general}---, since the two sets are complete, one can expand one set in terms of the other. From now on we concentrate on the outgoing modes ($\rightarrow$). We get
\be
u^{\ell m}_{\rightarrow \omega}=\int d\omega' \,\Big( \alpha^{\ell m\omega}_{\ell' m'\omega'}  \, u^{\ell' m' M}_{\rightarrow \omega'} +\beta^{\ell m\omega}_{\ell' m'\omega'} \bar u^{\ell' m' M}_{\rightarrow \omega'}\Big)\,,
\ee
where the $\alpha^{\ell m\omega}_{\ell' m'\omega'}$ and $\beta^{\ell m\omega}_{\ell' m'\omega'}$ are called Bogoliubov coefficients. Taking into account the orthonormality conditions \eqref{eq:orto}-\eqref{eq:ortoM} we get
\be
\alpha^{\ell m\omega}_{\ell' m'\omega'} = (u^{\ell' m' M}_{\rightarrow \omega'},u^{\ell m}_{\rightarrow \omega})\quad ,\quad
\beta^{\ell m\omega}_{\ell' m'\omega'}= - ( \bar u^{\ell' m' M}_{\rightarrow \omega'},u^{\ell m}_{\rightarrow \omega})\,,
\ee
and
\be\label{normBC}
\sum_{\ell'\in \N}\sum_{m'=-\ell'}^{\ell'} \int d\omega'\, \Big(\alpha^{\ell m\omega}_{\ell' m'\omega'} \alpha^{\ell' m'\omega'}_{\ell'' m''\omega''}  - \beta^{\ell m\omega}_{\ell' m'\omega'}  \bar \beta^{\ell' m'\omega'}_{\ell'' m''\omega''} \Big) = \delta (\omega ,\omega'')\,.
\ee
Moreover, defining the particle number operator for the mode $(\ell, m, \omega)$ in the $u^{\ell m}_{\rightarrow \omega}$-expansion in the usual form $N^{\ell m}_{\rightarrow \omega}=a^{\ell m \dagger}_{\rightarrow \omega} a^{\ell m}_{\rightarrow \omega} $, its expectation value on the Minkowski vacuum can generically be written as
\be
\bra{0} N^{\ell m}_{\rightarrow \omega} \ket{0} = \sum_{\ell'\in \N}\sum_{m'=-\ell'}^{\ell'}  \int d\omega' \, |\beta^{\ell m \omega}_{\ell' m' \omega'}|^2
\ee
This object is what we are mainly interested in. It tells us the expectation value of the number of excitations defined with respect to the conformal vacuum $\ket{0}$ that are present in the Minkowski quantum vacuum $\ket{0}_M$. The remarkable fact is that the computation of such object mimics exactly the one for the Hawking's particle production by a collapsing black hole.

Let us choose $\scri^+$ as the hypersurface over which we perform the integral for the computation of scalar products at least for the outgoing modes. $\scri^-$ would be the choice for the ingoing ones. In order to be able to use the near horizon approximate solutions \eqref{modes}, we introduce a complete set of outgoing wave packets on $\scri^+$ localized in retarded time $\bar u$ and near the horizon $\bar u\to+ \infty$ \cite{hawking1975}; see also \cite{fabbri2005modeling}. Concretely,  
\be\label{packets}
u^{\ell m; jn} = \frac{1}{\sqrt{\epsilon}} \int_{j\epsilon}^{(j+1)\epsilon} d\omega \, e^{2\pi i \omega n/\epsilon}\, u^{\ell m}_{\rightarrow \omega}
\ee
with integers $j \geq 0$, $n$, and where \be\label{outmodes}
u^{\ell m}_{\rightarrow \omega}=\frac{\sqrt\D}{ \sqrt{\omega}}\, \frac{e^{-i\omega \bar{u}}}{r} Y^{\ell m}(\vartheta, \varphi)\,.
\ee
 The wave packets $u^{\ell m; jn}$ are peaked around $\bar{u} \simeq 2\pi n/\epsilon$ with width $2\pi/\epsilon$. When $\epsilon$ is small, the wave packet is narrowly peaked about $\omega \simeq \omega_j = j\epsilon$ and localised near the horizon. 
 The facts that, due to spherical symmetry, the modes \eqref{modes} and \eqref{Mmodes} have exactly the same angular dependence, together with the fact that, in the region where the wave packets are picked, the behaviour in $\bar u$ and $u$ is independent of $\ell$, tells us that particle creation will be the same in all angular modes.
 
The surface element of $\scri^+$ is given by $d\Sigma^a = r^2 du\, dS^2 \delta^a_u $. The Bogoliubov coefficients of interest can therefore we written as
\be
\begin{split}
\beta^{\ell m ; jn}_{\ell' m' \omega'} &= - ( \bar u^{\ell' m' M}_{\rightarrow \omega'},u^{\ell m; jn})\\
&= i \int_{\scri^+} du\, dS^2\, r^2 \Big(u^{\ell m; jn}\partial_u u^{\ell' m' M}_{\rightarrow \omega'} -u^{\ell' m' M}_{\rightarrow \omega'} \partial_u u^{\ell m; jn}\Big)\,.
\end{split}
\ee
Since the wave packets vanish for $u \to -\infty$ and for $u>u_-$, we can integrate by part finding
\be
\beta^{\ell, m; jn}_{\ell' m' \omega'} = 2i \int_{-\infty}^{u_-} du\, dS^2\, r^2 \,u^{\ell m; jn}\partial_u u^{\ell' m' M}_{\rightarrow \omega'} \,.
\ee
We now need to insert in the equation the explicit form of the modes \eqref{packets}-\eqref{outmodes} and the outgoing part of the Minkowskian ones \eqref{Mmodes}. However, before doing that, let us recall that we are working and are mainly interested in the near horizon limit $\bar u \to +\infty$. In this limit, the inverse of the relation \eqref{uv} between the conformal retarded time $\bar{u}$ and the Minkowski $u$ simplifies into
\be
\bar{u} = \frac{2}{\sqrt\D} \operatorname{arcoth}\left(-\frac{u}{u_-}\right) \simeq \frac{1}{\sqrt\D} \log\left(\frac{u-u_-}{2 u_-} \right)\,.
\ee
So we can write
\be
\beta^{\ell, m; jn}_{\ell' m' \omega'} = \frac{\sqrt\D\delta_{\ell,\ell'}\delta_{m,m'}}{2\pi  \sqrt\epsilon} \int_{-\infty}^{u_-} du\, \int_{j\epsilon}^{\epsilon} d\omega \, e^{2\pi i \omega n/\epsilon} \sqrt{\frac{\omega'}{\omega}}\,e^{-i\omega \frac{1}{\sqrt\D} \log\left(\frac{u-u_-}{2u_-}\right)-i\omega' u }\,.
\ee
Defining now $x = u_- - u$ we get
\be
\beta^{\ell, m; jn}_{\ell' m' \omega'} = \frac{\sqrt\D\delta_{\ell,\ell'}\delta_{m,m'}}{2\pi\sqrt\epsilon}e^{-i \omega' u_-} \int_{0}^{+\infty} dx\, \int_{j\epsilon}^{\epsilon} d\omega \, e^{2\pi i \omega n/\epsilon} \sqrt{\frac{\omega'}{\omega}}\,e^{-i\omega \frac{1}{\sqrt\D} \log\left(\frac{x}{2u_-}\right) + i\omega' x }\,.
\ee
The integral over the frequency can be performed considering that $\omega$ varies in a small interval around $\omega_j$
\be
\begin{split}
\beta^{\ell, m; jn}_{\ell' m' \omega'} &= \frac{\sqrt\D\delta_{\ell,\ell'}\delta_{m,m'}}{\pi  \sqrt\epsilon} e^{-i \omega' u_-} \sqrt{\frac{\omega'}{\omega_j}}\int_{0}^{+\infty} dx\, e^{+i\omega' x} \,\frac{\sin (\epsilon L/2)}{L}\,e^{i L \omega_j}\\
&=\frac{\sqrt\D\delta_{\ell,\ell'}\delta_{m,m'}}{\pi  \sqrt\epsilon} e^{-i \omega' u_-} \sqrt{\frac{\omega'}{\omega_j}} \;I(\omega')\,,
\end{split}
\ee
where we have defined
\be
L(x) = \frac{2\pi n}{\epsilon}-\frac{1}{\sqrt\D}\log\left(\frac{x}{2u_-}\right)
\ee
and $I(\omega')$ as the integral over $x$. The computation of $\alpha_{jn,\omega'}$ gives a similar result
\be
\begin{split}
\alpha^{\ell, m; jn}_{\ell' m' \omega'}  &= \frac{\sqrt\D\delta_{\ell,\ell'}\delta_{m,m'}}{\pi  \sqrt\epsilon}e^{i \omega' u_-} \sqrt{\frac{\omega'}{\omega_j}}\int_{0}^{+\infty} dx\, e^{-i\omega' x} \,\frac{\sin (\epsilon L/2)}{L}\,e^{i L \omega_j}\\
&=\frac{\sqrt\D\delta_{\ell,\ell'}\delta_{m,m'}}{\pi  \sqrt\epsilon} e^{i \omega' u_-} \sqrt{\frac{\omega'}{\omega_j}} \;I(-\omega')\,.
\end{split}
\ee
Apart from different constants, these objects coincide with the ones defined in \cite{fabbri2005modeling}, and therefore can be solved using exactly the same techniques and procedure. We refer to the book for details and we give here only the final result. 

The important result is that the relation between $\alpha^{\ell, m; jn}_{\ell' m' \omega'} $ and $\beta^{\ell, m; jn}_{\ell' m' \omega'}$ comes out to be
\be
|\beta^{\ell, m; jn}_{\ell' m' \omega'}| = e^{-\frac{\pi \omega_j}{\sqrt\D}}\;|\alpha^{\ell, m; jn}_{\ell' m' \omega'} |\,.
\ee
Inserting this into Eq.~\eqref{normBC}, one can write
\be
-\left[1-\exp\left(\frac{2\pi \omega_j}{\sqrt\D} \right)\right] \sum_{\ell'\in \N}\sum_{m'=-\ell'}^{\ell'}   \int_0^{+\infty} d\omega'|\beta^{\ell, m; jn}_{\ell' m' \omega'}|^2 = 1
\ee
and therefore
\be\label{planck}
\bra{0} N^{\ell m}_{\omega_j} \ket{0}  = \frac{1}{\exp\left(\frac{2\pi \omega_j}{\sqrt\D} \right)-1}\,.
\ee
The above expression coincides with the Planck distribution of thermal radiation at the temperature
\be\label{temp}
T = \frac{\sqrt\D}{2\pi}\,.
\ee
To relate this result to the first law, Eq.~\eqref{firstLaw}, it is enough to notice that the explicit value of the conserved quantity $\s$ in our case is
\be
\s = \sqrt{\Delta}\,.
\ee
We have shown what we anticipated at the very beginning of this section: the light cone $u=u_-$ is seen as a (conformal) horizon with an associated temperature $T$ given by expression \eqref{temperature}.

The first law can therefore be rewritten as
\be
\delta M = T\delta S+\delta M_{\infty}\,.
\ee
with
\be
T = \frac{\s}{2\pi} \ \ \ \ \ \ {\rm and} \ \ \ \ \ \ \delta S= \frac{\delta A}{4}\,.
\ee
The laws of light cone thermodynamics are now not simply a mere analogy, but they acquire a precise semiclassical thermodynamical sense, which is better discussed in the following subsection. This is, to our knowledge, the first precise implementation of the idea \cite{sultana2004conformal,dyer1979conformal} that the quantity $\s$ should play the role of temperature for conformal Killing horizons.

 \subsection{On the meaning of conformal energy and temperature}\label{sec:energy}

In asymptotically flat stationary spacetimes, the time translational Killing field can be normalized at infinity in order to give the analogue of \eqref{M} the physical interpretation of energy as seen from infinity. On the other hand in our case the vector field $\xi$ is normalized only on the observer sphere $r=r_{\va O}$ and $t=0$. Thus $M$ has not the usual physical meaning for any observer in Minkowski spacetime. Nevertheless, for conformally invariant matter the mass $M$ as defined in (\ref{M}) is conformally invariant (see footnote~\ref{FOOT}). Using  \eqref{eq:confmetric}, and the fact that $\xi$ is actually a normalized Killing field of the static FRW metric, one can interpret $M$ as energy in the usual physical manner in that spacetime. 
This interpretation is compatible with the notion of frequency we used to compute the Planckian distribution \eqref{planck}. Indeed, the frequency $\omega$ is the one that would be measured by an observer moving along the Killing field $\partial_\tau$ in the static FRW space. 
For such notion of frequency $\omega$, the energy quanta $\varepsilon=\hbar \omega$ correspond to the same physical notion of energy that defines $M$.

Such interpretation carries over to its thermodynamical conjugate: the temperature. That is the reason why we call {\em conformal temperature} the temperature appearing in the first law. It carries the physical notion of temperature, namely the one measured by thermometers, only for observers in the FRW spacetime where $\xi$ is an actual time translational Killing field. In this way both energy and temperature have their usual interpretation in a spacetime that is conformally related to Minkowski.

\subsection{The Hartle-Hawking-like state}

Let us now define a new radial coordinate 
\be
R=\frac{\sqrt{\Delta}}{a} \exp(-\rho\sqrt{\Delta})\,.
\ee
The near horizon limit $\rho \to +\infty$ corresponds now to $R \to 0$. In these new coordinates, the metric \eqref{eq:confmetric} can be expanded around $R=0$ finding
\be\label{eq:metricR0}
ds_E^2=-R^2 d(\sqrt{\Delta}\tau)^2+dR^2+r_{\va H}^2 dS^2+O({R}{r^{-1}_H}\, dR^2, R r_H dS^2)\,,
\ee
where $O({R}{r^{-1}_H}\, dR^2, R r_H dS^2)$ denotes subleading terms of each component of the metric that do not change the nature of the apparent singularity present at $R=0$. Notice that the leading order of the local metric and the topological structure at the point $r=r_{\va H}$ are exactly the same as the one in the Reissner-Nordstrom metric, Eq.~\eqref{nhl}.

Moreover, the metric \eqref{eq:confmetric} can be continued analytically to imaginary conformal Killing time by sending $\tau\to- i\tau_E$. As for the case of static black holes \cite{wald2010general}, the result is a real Euclidean metric, explicitly given by
\be\label{metricEuclidean}
ds_E^2 = \Omega_{ E}^2 \left(d\tau_E^2 + d\rho^2 +\Delta^{-1} \sinh^2(\rho \sqrt{\Delta})dS^2\right)
\ee
with
\be
\Omega_{E} = \frac{\Delta/2a}{\cosh(\rho \sqrt{\Delta})-\cos(\tau_E \sqrt{\Delta})}.
\ee
Defining again the new coordinate $R$ and carrying out the limit to $R=0$, which corresponds to the Euclidean analogue of the horizon, we find the Euclidean version of \eqref{eq:metricR0}
\be
ds_E^2=R^2 d(\sqrt{\Delta}\tau_E)^2+dR^2+r_{\va H}^2 dS^2+O({R}{r^{-1}_H}\, dR^2, R r_H dS^2)\,.
\ee
The coordinate singularity at $R=0$ can be resolved by defining new coordinates $X=R \cos(\sqrt{\Delta }\tau_E)$ and $Y=R\sin(\sqrt{\Delta} \tau_E)$. In order to avoid conical singularities one must identify $\tau_E$ with a periodic coordinate such that
\be\label{periodictime}
0\le \tau_E \sqrt{\Delta}\le 2\pi\,.
\ee
This removes the apparent singularity by replacing the first two terms in the previous metric by the regular $dX^2+dY^2$ transversal metric. This periodicity in time is what is used in the black hole case to suggest the existence of a state---known as the Hartle-Hawking state---of thermal equilibrium of any quantum field at a temperature given by
\be\label{HHtemp}
T=\hbar \frac{\sqrt{\Delta}}{2\pi}\,,
\ee
which coincides with the one found in the previous section, Eq.~\eqref{temp}.
This tells us that the Minkowski vacuum can be regarded as the Hartle-Hawking type of vacuum of Region II for conformally invariant theories. 

Indeed, instead of using the near horizon approximation (and wave packets peaked there) in the previous section, one could in principle compute the Bogoliubov coefficients exactly between the Minkowski and conformal Fock spaces. 
This should lead to the conclusion that Minkowski vacuum is everywhere a thermal state with temperature \eqref{HHtemp}, as suggested by the previous analysis. As such computation might be rather involved and in view of keeping the presentation as simple as possible, one can find additional evidence for this  by computing the expectation value of the normal ordered stress energy ``tensor'' $:T_{ab}:$. The quotation marks on the word tensor are because the object $:T_{ab}:$ does not transform as a tensor under a coordinate transformation and cannot be interpreted physically as real. In fact the physical and covariant energy momentum tensor has vanishing expectation value in the Minkowski vacuum \cite{Wald:1995yp}.  

However,  $:T_{ab}:$ can be interpreted as encoding the particle content of the Minkowski vacuum as seen from the perspective of the MCKF that are of interest in our analysis. For conformal fields it can be analytically computed  in the $s$-wave approximation $\ell=0$ which reduces the calculation to an effective 2-dimensional system. The 2-dimensional $:T^{\va (2)}_{ab}:$ can be  explicitly evaluated via the Virasoro anomaly \cite{birrelldavies:QFTCST,fabbri2005modeling}. Given two sets of double null coordinates, like the two we have $(u,v)$ and $(\bar{u},\bar{v})$, $:T^{\va (2)}_{ab}:$ transforms as
\be
\begin{split}
:T^{\va (2)}_{\bar{u}\bar{u}}: &= \left(\frac{du}{d\bar{u}}\right)^2 :T^{\va (2)}_{uu}: -\frac{\hbar}{24\pi} \{u,\bar{u}\}\\
:T^{\va (2)}_{\bar{v}\bar{v}}: &= \left(\frac{dv}{d\bar{v}}\right)^2 :T^{\va (2)}_{uu}: -\frac{\hbar}{24\pi} \{v,\bar{v}\}
\end{split}
\ee
where 
\be
\{x,y\} =\frac{\dddot x}{\dot x} -\frac{3}{2} \left( \frac{\ddot x}{\dot x}\right)^2
\ee
is the Schwarzian derivative with dot representing $d/dy$. It is therefore simple to evaluate the expectation value of this object on the Minkowski vacuum $\ket{0}_M$ in our case. Since $_M\bra{0}:T_{ab}:\ket{0}_M=0$, we simply have
\be
\begin{split}
_M\bra{0}:T^{(\va 2)}_{\bar{u}\bar{u}}:\ket{0}_M = -\frac{\hbar}{24\pi } \{u,\bar{u}\} &= \frac{\hbar \Delta}{48 \pi } \\
_M\bra{0}:T^{\va (2)}_{\bar{v}\bar{v}}:\ket{0}_M =  -\frac{\hbar}{24\pi } \{v,\bar{v}\} &= \frac{\hbar \Delta}{48 \pi } \,.
\end{split}
\ee
The result indicates that the Minkowski state produce a constant ingoing and outgoing thermal bath at the temperature \eqref{temp} everywhere in Region II, which is what we expected from a thermal equilibrium state. The near horizon approximation in the computation of the previous section simplifies the relation between the two sets of double null coordinates making the computation analytically simpler, but as discussed above, it should give the same result everywhere in Region II. As mentioned above, the expectation value of the covariant stress energy tensor does not coincide with the normal ordered one. The former is simply vanishing in this case $_M\bra{0}T_{ab}\ket{0}_M=0$. 

\begin{figure}[t]
\center
\includegraphics[height=10cm]{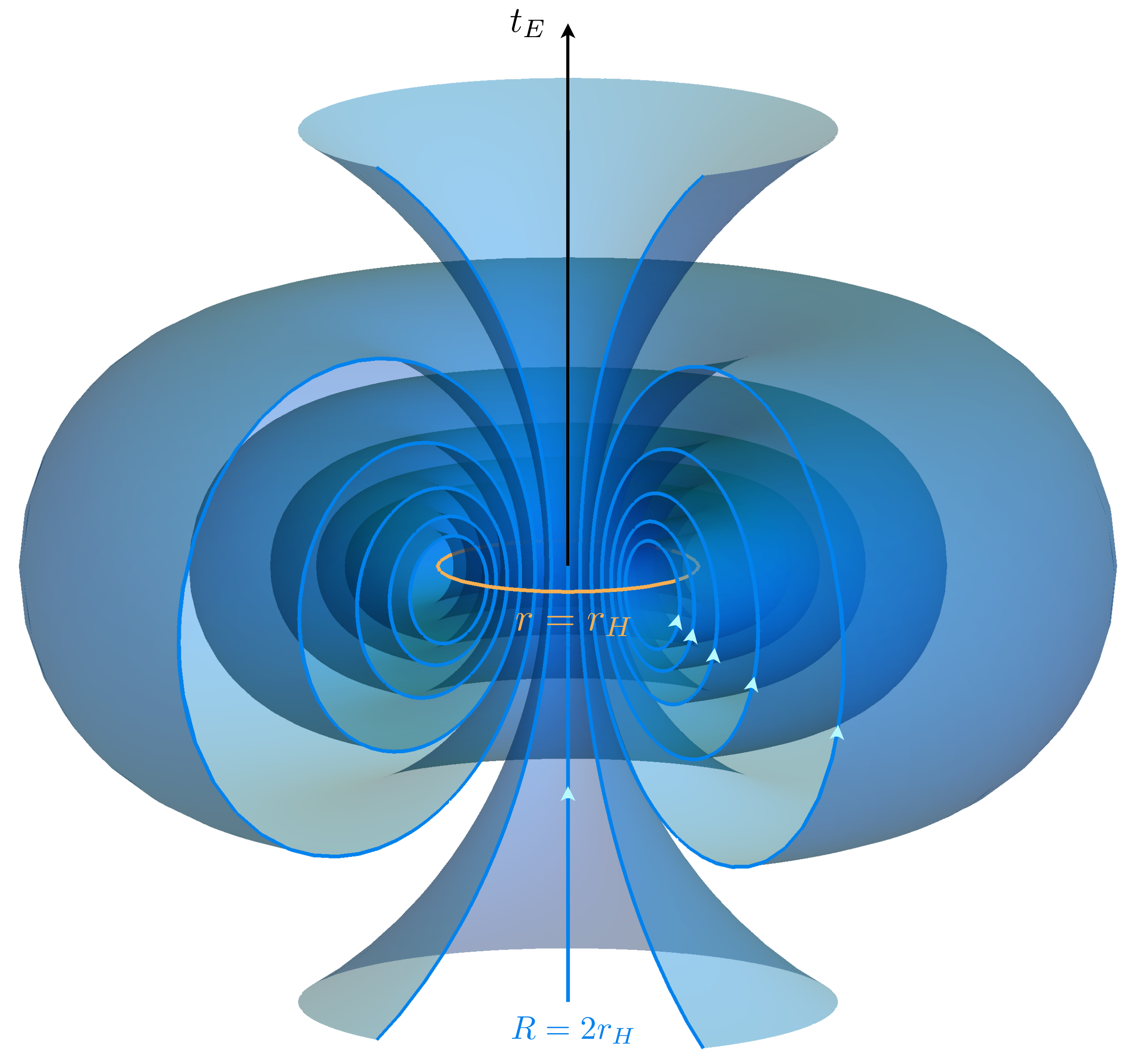}
\caption{Three dimensional representation of the flow of the conformal Killing field in the Euclidean spacetime $\R^4$. The orbits in this one-dimension-less representation are non-concentric tori around the bifurcate sphere $r=r_H$---here represented as a circle. They degenerate into the $t_E$ axis for $R=2r_H$.}
\label{fig:Euclidean}
\end{figure}

As a final remark, let us get more insight into the geometry of the Euclidean continuation \eqref{metricEuclidean} by writing the coordinate transformation to the flat Euclidean coordinates $(t_E, r,\vartheta, \varphi)$ covering $\R^4$. Defining the angular coordinate $\alpha_E\equiv \tau_E \sqrt{\Delta}$ one finds
\be
\label{eq:cootranEu}
\begin{split}
t_E &=\frac{R \sin(\alpha_E)}{1-\frac{R}{2r_{H}} \cos(\alpha_E) +\frac{R^2}{4r^2_{H}}}\\
r &= r_H \frac{1-\frac{R^2}{4r^2_{H}}}{1-\frac{R}{2r_{H}} \cos(\alpha_E) +\frac{R^2}{4r^2_{H}}}\,.
\end{split}
\ee
The bifurcate sphere in the Euclidean continuation corresponds to the sphere $r=r_H$ at $t_E=0$. The orbits of the Wick rotated radial conformal Killing field are orbits of the radial conformal Killing field of $\R^4$ with fixed points given by the Euclidean shining sphere. These orbits correspond, on the $(t_E, r)$ plane, to close loops around the bifurcate sphere, which degenerate into the $r=0$ line (the Euclidean $t_E$-axis) for $R=2r_H$; see Figure~\ref{fig:Euclidean}. The  coordinates $(\tau_E, R, \vartheta, \varphi)$ become singular there. The qualitative features of the Euclidean geometry of the MCKF is just analogous to that of the stationarity Killing field in the Euclidean RN solutions.

\section{Discussion}

We have studied in detail the properties of radial Conformal Killing Fields in Minkowski spacetime and showed that they present in many respects a natural analogue of black holes in curved spacetimes. The global properties of radial MCKFs mimic exactly the causal separation of events in the spacetime of static black holes, i.e. those in the Reissner-Nordstrom family; see Figure \ref{fig:penrose}. Event Killing horizons in the latter are replaced by conformal Killing horizons in the former. The extremal limit maintains the correspondence. 

Linear perturbations of flat Minkowski spacetime in terms of conformally invariant matter models, allow us to consider and prove suitable analogues of the laws of black hole mechanics. When quantum effects are considered, thermal properties make the classical mechanical laws amenable to a suitable thermodynamical interpretation, where entropy variations are equal to $1/4$ of the conformal area changes of the horizon in Planck units. The near horizon and near bifurcate surface features of the geometry of the radial MCKF have the same structure of the stationarity Killing field for static black holes. The Minkowski vacuum state is the analogue of the Hartle-Hawking thermal state from the particle interpretation that is natural to the MCKF. 

This work represents another simple setting where the relationship between thermality, gravity and geometry is manifest in the semiclassical framework. It gives a simple and complete example in which thermal properties analogous to those of black holes are manifest in flat spacetime. It improves the standard analogy given by the study of gravity perturbations and quantum field theory of the Rindler wedge.
On a more speculative perspective, we think that even when the interpretation of temperature, energy, and area entering the thermodynamical relations is subtle, this simple example could shed some light into a more fundamental description of the link between black hole entropy and (quantum) geometry. But this is something we will investigate in the future. 

\section{Acknowledgments}

We acknowledge the anonymous referee for the constructive interaction that really improved this work. We thank Thibaut Josset for useful discussions, and Diego Forni for exchanges in the framework of an early version of this project. We would also like to thank the hospitality of the relativity group at the Fa.M.A.F. (Universidad Nacional de C\'ordoba) where an important part of this work was done.

\appendix
\section{Coordinate  Transformations}\label{app:coordinate}
The radial Conformal Killing Field in Minkowski spacetime $\xi$ naturally divides the space in six regions. For each of these regions, there exists a coordinate transformation $(t,r,\vartheta,\varphi) \to (\tau,\rho,\vartheta,\varphi)$ adapted to the MCKF in the sense that $\xi(\tau)=-1$. In this Appendix, we write down such transformations explicitly. 

\subsection*{The non-extremal case $\Delta\not=0$.}
\noindent
\textbf{Region I} {(\em the diamond)}{\bf.} The coordinate transformation is given by \cite{HalILQGS}
\be
\begin{split}
t &= \frac{\sqrt{\Delta}}{2a} \frac{\sinh(\tau \sqrt{\Delta})}{\cosh(\rho \sqrt{\Delta}) + \cosh(\tau \sqrt{\Delta})}\\
r &=\frac{\sqrt{\Delta}}{2a} \frac{\sinh(\rho \sqrt{\Delta})}{\cosh(\rho \sqrt{\Delta}) + \cosh(\tau \sqrt{\Delta})}
\end{split}
\ee
with $-\infty< \tau < + \infty$ and $0 < \rho < +\infty$.

\be
\begin{split}
v &= t+r = \frac{\sqrt{\Delta}}{2a} \tanh \frac{\bar{v}\sqrt{\Delta}}{2} \\ 
u &= t-r = \frac{\sqrt{\Delta}}{2a} \tanh \frac{\bar{u}\sqrt{\Delta}}{2} \, .
\end{split}
\ee
where we have defined the null coordinates $\bar{v}=\tau + \rho$ and $\bar{u}=\tau-\rho$.
The Minkowski metric \eqref{eq:min}  in the new coordinates reads
\be
ds^2 = \Omega_{\rm I}^2 \left(-d\tau^2 + d\rho^2 +\Delta^{-1} \sinh^2(\rho \sqrt{\Delta})dS^2\right)
\ee
with
\be
\Omega_{\rm I} = \frac{\Delta/2a}{\cosh(\rho \sqrt{\Delta})+\cosh(\tau \sqrt{\Delta})}.
\ee
\noindent
\textbf{Regions II} {\em (the causal complement of the diamond)}{\bf , III and IV.} Region II, III and IV can be described by the same coordinate transformation given by \cite{HalILQGS}
\be
\label{eq:cootranApp}
\begin{split}
t &=\frac{\sqrt{\Delta}}{2a} \frac{\sinh(\tau \sqrt{\Delta})}{\cosh(\rho \sqrt{\Delta}) - \cosh(\tau \sqrt{\Delta})}\\
r &=\frac{\sqrt{\Delta}}{2a} \frac{\sinh(\rho \sqrt{\Delta})}{\cosh(\rho \sqrt{\Delta}) - \cosh(\tau \sqrt{\Delta})}\,,
\end{split}
\ee
with $-\infty< \tau < + \infty$ and $0 \leq \rho < +\infty$. Region II is now given by the restriction $|\tau|<\rho$, Region III by $\tau > 0$ and $\tau > \rho$, while Region IV by $\tau < 0$ and $|\tau| > \rho$. In this case we have
\be\label{eq:uvapp}
\begin{split}
v &= t+r = -\frac{\sqrt{\Delta}}{2a} \coth\frac{\bar{u}\sqrt{\Delta}}{2}\\ 
u &= t-r = -\frac{\sqrt{\Delta}}{2a} \coth\frac{\bar{v}\sqrt{\Delta}}{2}\,.
\end{split}
\ee
The metric \eqref{eq:min} is now
\be
ds^2 = \Omega_{\rm II}^2 \left(-d\tau^2 + d\rho^2 +\Delta^{-1} \sinh^2(\rho \sqrt{\Delta})dS^2\right)
\ee
with
\be
\Omega_{\rm II} = \frac{\Delta/2a}{\cosh(\rho \sqrt{\Delta})-\cosh(\tau \sqrt{\Delta})}.
\ee
For Region II, given the above mentioned restrictions on the coordinate, we have $\bar{u} \in (-\infty,0)$ and $\bar{v} \in (0,+\infty)$. This is the transformation used in Section~\ref{HRCT}.

\noindent
\textbf{Region V.}
In the upper of the two regions where $\xi$ is spacelike, the coordinate transformation can be found to be
\be
\begin{split}
t &= \frac{\sqrt{\Delta}}{2a} \frac{\cosh(\tau \sqrt{\Delta})}{\sinh(\rho \sqrt{\Delta}) + \sinh(\tau \sqrt{\Delta})}\\
r &=\frac{\sqrt{\Delta}}{2a} \frac{\cosh(\rho \sqrt{\Delta})}{\sinh(\rho \sqrt{\Delta}) + \sinh(\tau \sqrt{\Delta})}
\end{split}
\ee
with $0< \tau < + \infty$ and $0 < \rho < +\infty$.
The double null coordinates are here given by
\be
\begin{split}
v &= t+r = \frac{\sqrt{\Delta}}{2a} \coth \frac{\bar{v}\sqrt{\Delta}}{2} \\ 
u &= t-r = \frac{\sqrt{\Delta}}{2a} \tanh \frac{\bar{u}\sqrt{\Delta}}{2} \, .
\end{split}
\ee

\noindent
\textbf{Region VI.}
Finally, for Region VI we have
\be
\begin{split}
t &= \frac{\sqrt{\Delta}}{2a} \frac{\cosh(\tau \sqrt{\Delta})}{\sinh(\rho \sqrt{\Delta}) - \sinh(\tau \sqrt{\Delta})}\\
r &=\frac{\sqrt{\Delta}}{2a} \frac{\cosh(\rho \sqrt{\Delta})}{\sinh(\rho \sqrt{\Delta}) - \sinh(\tau \sqrt{\Delta})}
\end{split}
\ee
with $-\infty< \tau < 0$ and $0 < \rho < +\infty$. This gives
\be
\begin{split}
v &= t+r = -\frac{\sqrt{\Delta}}{2a} \tanh \frac{\bar{u}\sqrt{\Delta}}{2} \\ 
u &= t-r = -\frac{\sqrt{\Delta}}{2a} \coth \frac{\bar{v}\sqrt{\Delta}}{2} \, .
\end{split}
\ee
In both last two cases, the metric \eqref{eq:min} becomes
\be
ds^2 = \Omega_{V/VI}^2 \left(-d\tau^2 + d\rho^2 +\Delta^{-1} \cosh^2(\rho \sqrt{\Delta})dS^2\right)
\ee
where, for Region V
\be
\Omega_{\rm V} = \frac{\Delta/2a}{\sinh(\rho \sqrt{\Delta})+\sinh(\tau \sqrt{\Delta})}\,,
\ee
and for Region VI
\be
\Omega_{\rm VI} = \frac{\Delta/2a}{\sinh(\rho \sqrt{\Delta})-\sinh(\tau \sqrt{\Delta})}\,.
\ee

\subsection*{The extremal case $\Delta = 0$}

In the $\Delta=0$ case, we have only Region II, III and IV and $\xi$ is everywhere timelike. The coordinate transformation in this extremal case can be obtained from the previous one by taking the limit $\Delta\to 0$ in all expressions. The result is
\be
\begin{split}
t=&\frac{\tau }{a \left(\tau ^2-\rho ^2\right)}\\
r=&\frac{\rho }{a \left(\rho ^2-\tau ^2\right)}
\end{split}
\ee
with $-\infty< \tau < + \infty$ and $0 \leq \rho < +\infty$. Region II is now given by the restriction $|\tau|<\rho$, Region III by $\tau > 0$ and $\tau > \rho$, while Region IV by $\tau < 0$ and $|\tau| > \rho$. In this case we have
\be
\begin{split}
v &= t+r = \frac{1}{a \bar v} \\ 
u &= t-r =  \frac{1}{a \bar u},
\end{split}
\ee
where, given the above mentioned restrictions on the coordinate, we have $\bar{u} \in (-\infty,0)$ and $\bar{v} \in (0,+\infty)$.
The Minkowski metric  in the new coordinates reads
\be
ds^2 = \Omega_{\rm ext}^2 \left(-d\tau^2 + d\rho^2 +\rho^2 dS^2\right)
\ee
with
\be
\Omega_{\rm ext} = \frac{\rho }{a \left(\rho ^2-\tau ^2\right)}.
\ee
This coincides with Eq.~(12) in \cite{RCKF}.

\subsection*{Near bifurcate sphere approximation}
In the non-extremal case, the bifurcate sphere is located at $\rho \to +\infty$ and $\tau=0$. Eq.~\eqref{eq:cootran} can therefore be expanded in the approximation $\rho >> 1/\sqrt{\Delta}$. This gives a Rindler-like coordinate transformation
\be
\begin{split}
t \sim -\sqrt{\frac{c}{a}} \;e^{-\rho\sqrt{\Delta}} \cosh(\tau \sqrt{\Delta})\\
r \sim \sqrt{\frac{c}{a}} \;e^{-\rho\sqrt{\Delta}} \sinh(\tau \sqrt{\Delta})
\end{split}
\ee
with the would-be proper distance given by $D = \sqrt{c/a} \;e^{-\rho\sqrt{\Delta}}$. The above approximation is inconsistent in the case $\Delta = 0$.

\section{Static FRW Spacetime and Region II}\label{app:FRW}
\begin{figure}[t]
\center
\includegraphics[height=10cm]{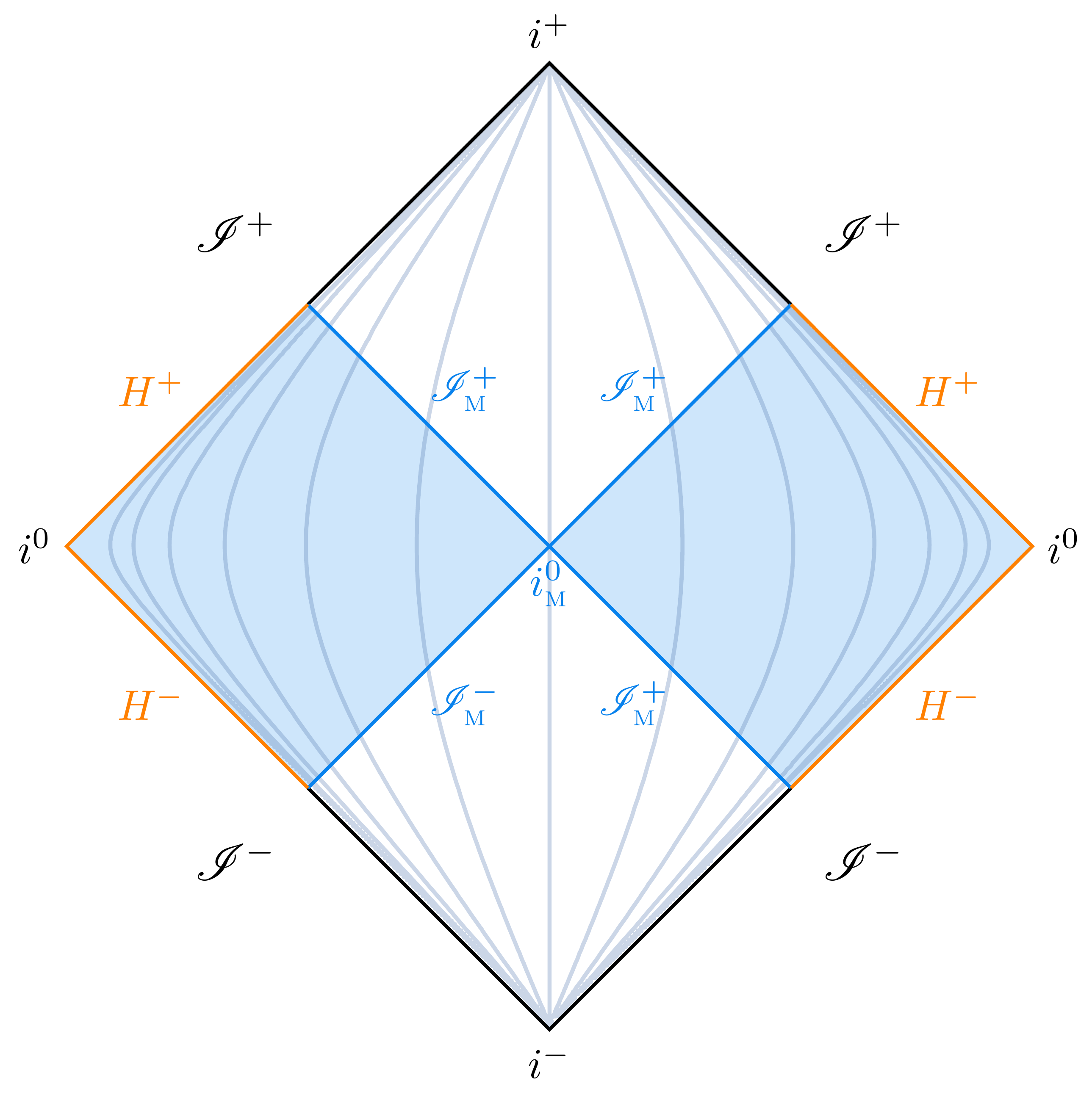}
\caption{The Penrose diagram for the static FRW spacetime of Eq.~\eqref{eq:FRWmetric}. The shaded region is the one conformally related to Region II in Minkowski spacetime. Its boundaries are in correspondence with pieces of Minkowskian future and past null infinities $\scri^\pm_M$, as well as with the bifurcate conformal Killing horizon $H^+ \cup H^-$. The light grey hyperbolas are radial flow lines of the field $\partial_\tau$, or in another words $\rho = const$ lines.}
\label{fig:FRW}
\end{figure}

The coordinate transformations above show that the Regions I to IV in Minkowski spacetime are conformally related to pieces of a static FRW spacetime with negative spatial curvature $k=-|\Delta|$.
This fact was used in the computation of Bogoliubov coefficients in Section~\ref{HRCT}. In this Appendix we give some more details on the geometry of static FRW spacetime and its relation with Region II. The static FRW spacetime is a solution to the Einstein equation with zero cosmological constant and the energy stress tensor of a perfect fluid satisfying the state equation \cite{hawking1973large}
\be
\mu = - 3 p\,.
\ee
Here $\mu$ and $p$ are the energy density and pressure of the fluid respectively. The metric takes the form
\be\label{eq:FRWmetric}
ds^2 = -d\tau^2 + d\rho^2 +\Delta^{-1} \sinh^2(\rho \sqrt{\Delta})dS^2\,,
\ee
with $-\infty< \tau < + \infty$ and $0 \leq \rho < +\infty$. As shown for example in \cite{Candelas:1978gf,hawking1973large}, there exists a coordinate transformation that conformally maps this space into the Einstein static universe. This transformation allows to draw the Penrose diagram for the static FRW spacetime, which results in a diamond shaped diagram depicted in Figure~\ref{fig:FRW} \footnote{In the cited references \cite{Candelas:1978gf,hawking1973large}, however, they consider the non-static FRW spacetime with zero cosmological constant $\Lambda = 0$ and zero pressure $p=0$. The resulting Penrose diagram is therefore slightly different, being only the upper triangle of the whole diamond of Figure~\ref{fig:FRW}, with a ``big bang singularity'' for $\tau = 0$.}. 
The structure is very similar to the one of Minkowski, with past and future null infinities. There is however one main difference: here spacial infinity $i^0$ is a sphere and not a point as in the Minkowski case.

Region II is (conformally) given by the restriction $|\tau|<\rho$. This corresponds to the region outside the light cone shining from the origin; this is the shaded region in Figure~\ref{fig:FRW}. From Eq.~\eqref{eq:uvapp}, Minkowski future null infinity $v \to +\infty$ is mapped into the future light cone $\bar u = 0$. Analogously, Minkowski past null infinity $u \to - \infty$ is given by the past light cone $\bar v = 0$. In the same way, the future horizon $H^+$ located at $u = u_-$ is given by $\bar v \to +\infty$, namely the piece of FRW $\scri^+$ given by $\bar u < 0$. The past horizon $v = v_+$ is, in a similar way, mapped into the piece of FRW $\scri^-$ given by $\bar v > 0$. Finally, Minkowskian spacial infinite $i^0_M$ is mapped into the point given by the origin, while the bifurcate sphere $r=r_H$ at $t=0$ is given by the FRW spacial infinity $i^0$.
In Figure~\ref{fig:FRW}, the flow lines of the Killing field $\partial_\tau$, namely $\rho = const.$ surfaces are also plotted. From there, the behaviour in Region II of the conformal Killing field $\xi$ depicted in Figure~\ref{fig:families} becomes clear. The flow lines of $\xi$ start their life on $\scri^-$ to end on $\scri^+$. 

The above discussion shows also that the complete set of solutions \eqref{modes} to the Klein-Gordon equation \eqref{eq:KG} is a good complete set of solutions also in our region of interest. The set considered, indeed, is regular everywhere, at the origin too, where otherwise there could have been a problem in our setting.

\section{Another conformal mapping of Minkowski}\label{AP2}

The previous map to a suitable FRW spacetime is useful for the calculations in Section \ref{HRCT}. However, it is not the best suited for the geometrical interpretation due to the fact that the horizon $H$ is mapped to infinity in the FRW spacetime. Here we construct a new conformal mapping of flat spacetime where $\xi$ becomes a Killing field, and the horizon is mapped to a genuine Killing horizon embedded in the bulk of the host spacetime. In Minkowski spacetime we have
\be
\sL_{\xi} \eta_{ab}=\frac{\psi}{2} \eta_{ab}
\ee
where explicit calculation yields
\be \psi= \nabla_a\xi^a=8at=\frac{4(u+v)}{r_{\va O}^2-r_{\va H}^2}.\ee
Under a conformal transformation $g_{ab}=\Omega^2 \eta_{ab}$ one has
\be
\sL_{\xi}g_{ab}=\sL_{\xi}(\Omega^2 \eta_{ab})=\left[\frac{\psi}{2} + 2 \xi(\log(\Omega))\right]g_{ab}.
\ee
Therefore, in the new spacetime $g_{ab}$, $\xi$ will be a Killing field iff \be {\psi}+ 4 \xi(\log(\Omega))=0.\ee This equation does not completely fix $\Omega$: if $\Omega$ is a solution, then $\Omega^{\prime}=\omega \Omega$ is also a solution as long as $\xi(\omega)=0$. Writing explicitly the previous equation  using (\ref{eq:CKFb0}) we get:
\be\label{ew}
 u+v + (u^2-r_{\va H}^2) \partial_u(\log(\Omega))+ (v^2-r_{\va H}^2) \partial_v(\log(\Omega))=0.
\ee
It is easy to find solutions of the previous equation by separation of variables. Assuming that we want to preserve spherical symmetry then we can write
$\Omega(u,v)=V(v)U(u)$ and the previous system becomes
\ba
 \notag  u + (u^2-r_{\va H}^2) \partial_u(\log(U))&=&-\lambda,\\
  v + (v^2-r_{\va H}^2) \partial_v(\log(V))&=&\lambda,
\ea
where $\lambda$ is an arbitrary constant. If we choose $\lambda=0$ then the solution is
\be
\Omega=\frac{\Omega_0}{\sqrt{(u^2-r_{\va H}^2)(v^2-r_{\va H}^2)}}.
\ee
By fixing the integration constant $\Omega_0=r_{\va O}^2-r_{\va H}^2$, the previous solution corresponds to the  one that maps to the FRW spacetime studied in the previous section. This can be checked by recalling from (\ref{eq:FRWmetric}) that the conformal factor mapping to the FRW spacetime is $1/\sqrt{-\xi\cdot \xi}$ and  using (\ref{eq:norm}). The Killing vector $\xi$ in the FRW metric is normalized everywhere.

An alternative solution, which does not send the horizon to infinity, is obtained by choosing $\lambda=r_{\va H}$ which yields
\be\label{OBH}
\Omega_{\va BH}=\frac{4r^2_{\va H}}{(u-r_{\va H})(v+r_{\va H})},
\ee
where we have chosen the integration constant so that $\Omega_{\va BH}=1$ at the bifurcate surface $u=-r_{\va H}$ and $v=r_{\va H}$. 
In the new metric 
\be
g_{ab}=\frac{16 r^4_{\va H}}{(u-r_{\va H})^2(v+r_{\va H})^2}\, \eta_{ab}, 
\ee
the null surfaces $u=-r_{\va H}$ and $v=r_{\va H}$ are Killing horizons with constant cross-sectional area $A=4\pi r_{\va H}^2$.  These surfaces have the same geometric properties as black hole horizons which justifies the subindex BH in (\ref{OBH}).  At the inner horizons $u=r_{\va H}$  and $v=-r_{\va H}$ the conformal factor diverges. Therefore, in contrast with the FRW mapping, only these  horizons are pushed to infinity. If $r_O\le \sqrt{5} r_H$, then the Killing field $\xi$ is normalized on a timelike surface outside the horizon where stationary observers measure time and energy in agreement with those in the FRW mapping \footnote{If one fixes $\Omega=1$ at the horizon and denotes $r_S$ the radius of the sphere defined by the intersection of the stationarity surface  $\xi\cdot\xi=-1$  and the $t=0$ hyperplane, then $r_H\le r_S\le \infty$ when $r_O$ moves in the interval $r_H\le r_O\le \sqrt{5} r_H$. }.  More details on these geometries can be found in \cite{myThesis}.  The previous conformal map plays a central role for the interpretation of the first law (\ref{firstLaw})  as discussed at the end of Section \ref{sec:firstlaw}.

%

\bibliographystyle{JHEP}
\bibliography{references}

\end{document}